\documentclass[12pt]{article}

\usepackage{amsfonts}   
\usepackage{amssymb}    
\usepackage{amsmath}

\usepackage{amsfonts} 
\DeclareMathSymbol{\shortminus}{\mathbin}{AMSa}{"39}

\usepackage{arxiv}
\usepackage{algorithm, algorithmic}

\usepackage[nocomma]{optidef}
\usepackage{mathtools}
\usepackage{empheq}
\usepackage{graphicx}
\numberwithin{equation}{section}

\usepackage[utf8]{inputenc} 
\usepackage[T1]{fontenc}    
\usepackage{hyperref}       
\usepackage{url}            
\usepackage{amsfonts}       
\usepackage{booktabs}       
\usepackage{nicefrac}       
\usepackage{microtype}      
\usepackage{lipsum}
\usepackage{graphicx}
\graphicspath{ {./images/} }
\usepackage{float}

\usepackage{caption}
\DeclareMathOperator{\minimize}{\mathbf{minimize}}
\DeclareMathOperator{\argmin}{\mathbf{arg-min}}
\DeclareMathOperator{\st}{\mathrm{s.t.}}

\newcommand{\bigbs}{\!\!\!\!}
\newcommand{\dt}{\,\texttt{dt}}
\newcommand{\F}{\mathcal{F}}

\newcommand{\blam}{b_\lambda}
\newcommand{\blamA}{b_{\lambda,1}}
\newcommand{\blamB}{b_{\lambda,2}}
\newcommand{\blamG}{b_{\lambda, \gamma}}

\newcommand{\blamF}{b_{\lambda\F}}

\newcommand{\dotblam}{\dot{b}_{\lambda}}
\newcommand{\dotblamF}{\dot{b}_{\lambda \F}}

\newcommand{\Cost}{\text{Cost}}
\newcommand{\Costab}{\Cost(a, \blam)}
\newcommand{\Costba}{\Cost(\blam, a)}

\newcommand{\CostF}{\Cost_{\F}}
\newcommand{\CostFab}{\Cost_{\F,a}(\blam)}
\newcommand{\CostFba}{\Cost_{\F,\blam}(a)}

\newcommand{\Cons}{{\mathfrak{C}}}

\newcommand{\sumodd}{\sum_{\substack{n=1 \\ n\,\text{odd}}}^{N}}

\newcommand{\sumnmodd}{\sum_{\substack{n, m=1 \\ n+m\, \text{odd}}}^N}

\newcommand{\R}{\mathbf{R}} 

\newcommand{\T}{\mathcal{T}}

\newcommand{\acos}{\sum_{n=1}^N a_n n \pi \cos(n \pi t)}
\newcommand{\bcos}{\sum_{n=1}^N b_n n \pi \cos(n \pi t)}
\newcommand{\asin}{\sum_{n=1}^N a_n \sin(n \pi t)}
\newcommand{\bsin}{\sum_{n=1}^N b_n \sin(n \pi t)}

\newcommand{\intzo}{\int_0^1}

\newcommand{\intacos}{\sum_{n=1}^N \intzo a_n n \pi \cos(n \pi t)\dt}

\newcommand{\intasin}{\sum_{n=1}^N \intzo a_n \sin(n \pi t) \dt }

\newcommand{\sigkapplam}{{\kappa,\!\lambda,\!\sigma\!}}

\newcommand{\aeq}{a_{\texttt{\tiny eq}}}
\newcommand{\beq}{b_{\texttt{\tiny eq}}}

\newcommand{\ak}{a^{(k)}}
\newcommand{\bk}{\blam^{(k)}}

\usepackage{authblk}

\usepackage{arxiv}

\usepackage[english]{babel}

\usepackage{cleveref}
\crefname{equation}{Eq.}{Eqs.}
\Crefname{equation}{Equation}{Equations}

\usepackage{amsthm}

\theoremstyle{definition}

\newtheorem{prop}{Proposition}[section]

\usepackage[utf8]{inputenc} 
\usepackage[T1]{fontenc}    
\usepackage{hyperref}       
\usepackage{url}            
\usepackage{booktabs}       
\usepackage{amsfonts}       
\usepackage{nicefrac}       
\usepackage{microtype}      
\usepackage{lipsum}
\usepackage{graphicx}
\graphicspath{ {./images/} }

\usepackage{csquotes}

\begin{document}


\title{Position-building in competition with real-world constraints\thanks{This is version 2 of this paper and contains corrections to typos.}}

\author[1]{Neil A. Chriss}

\affil[1]{neil.chriss@gmail.com}

\maketitle


\begin{abstract}
This paper extends the optimal-trading framework developed \cite{arXiv:2409.03586v1} to compute optimal strategies with {\em real-world constraints}. The aim of the current paper, as with the previous, is to study trading in the context of {\em multi-player non-cooperative games}. While the former paper relies on methods from the calculus of variations and optimal strategies arise as the solution of partial differential equations, the current paper demonstrates that the entire framework may be re-framed as a quadratic programming problem and cast in this light constraints are readily incorporated into the calculation of optimal strategies. An added benefit is that two-trader equilibria may be calculated as the end-points of a dynamic process of traders forming repeated adjustments to each other's strategy.
\end{abstract}

\section{Introduction}
\label{sec:introduction}

    Almgren and Chriss developed a robust theory for trading portfolio transactions in {\em Optimal execution of portfolio transactions}\footnote{See \cite{almgren2001optimal} and \cite{almgren1997optimal}.}. That theory and the large body of work that followed is based on the trade-off between transaction costs and risk. The faster one trades, the greater the transaction costs but the more rapidly risk is reduced and vice-versa. A key feature of the framework is the {\em absence} of competition from other traders. By contrast \cite{arXiv:2409.03586v1} developed a framework for optimal trading that specifically {\em includes} competition between two or more traders in the same stock. The nature of the competition studied is {\em non-cooperative}, collusion is strictly prohibited.
    
    In the two-trader context, best-response and equilibrium are the central elements of the framework placing trading  in a game-theoretic context, specifically {\em non-cooperative games} where coordination between traders is strictly prohibited. A best-response strategy is a maximally cost-reducing strategy taking into account an adversary's strategy. Equilibrium is when two strategies are simultaneously the best-response to one another; \cite{arXiv:2409.03586v1} demonstrated that methods from the calculus of variation may be used to derive best-response strategies and two-trader equilibria. The paper additionally contains a definition of $n$-trader symmetric equilibria and explores how empirical aspects of transaction costs and competitive trading influence the structure of optimal strategies.

    A  limitation of \cite{arXiv:2409.03586v1} is that it is limited to {\em unconstrained} strategies. In real-world applications it is of considerable interest to find strategies constrained by practical considerations such as position and trading limits, including short-selling. This paper presents computational methods for finding best-response and equilibrium strategies with arbitrary constraints and in the case of two-trader equilibrium sheds additional light on the nature of equilibrium by studying that dynamic path to equilibrium. Therefore if \cite{arXiv:2409.03586v1} is about the theoretical foundations of {\em trading as game theory}, this paper concerns methods for efficient computation with real-world constraints. 

    In the immediate following sections we review the definition of trading strategies, best-response strategies and equilibrium. We then show we can approximate the relevant cost functions arbitrarily well by positive-definite quadratic functions and that constraints are linear. With this we are able to translate the minimizations achieved through variational methods to convex quadratic programs in a small number of parameters with linear constraints. 

    \subsection{Definitions and terminology}
    \label{sec:definitions}

    In this section we define the basic concepts in this paper and introduce standard terminology and notation. Key definitions will be introduced with in separate paragraphs beginning with bold face text.

    {\bf Position-building:} A position-building strategy seeks to acquire a {\em target quantity} of stock over a fixed period of time. In this context the motivation for acquiring a position is a {\em catalyst} that is expected to cause a {\em revaluation} of the stock.

    \vskip 10pt

    {\bf Competition:} When two or more traders are simultaneously trading the same stock at the same time they are {\em trading in competition.} The competition studied in in \cite{arXiv:2409.03586v1} and in this paper is {\em non-cooperative}, that is, traders  act independently and trading occurs without communication or coordination between traders. 

    \vskip 10pt

    {\bf Trading strategies:} A trading strategy is a description of {\em how} a stock is acquired over time; strategies are represented by twice-differentiable function from the unit interval to the real numbers. The unit interval represents time and the real numbers represent the  quantity of stock held at a given time. Trading strategies implicitly trade a specific but unnamed stock. For this paper all trading strategies start at zero and end at the target quantity. Because we will be dealing with two or more traders trading at once, we can always scale one of the traders target quantities to one, which we call a {\em unit strategy}. Strategies are denoted by the symbols $a$ or $b$. We use the symbol $\lambda$ for the target quantity of a Non-unit strategy, write $\blam$ for the strategy, where $b$ is the associated unit strategy and $b_\lambda$ is taken the function $\lambda b$. A feature and limitation of this paper and \cite{arXiv:2409.03586v1} is that the strategies considered are {\em static} throughout the course of trading. In this way, strategies may be treated as "plays" in games and standard concepts from game theory (see below) may be applied. 

    \vskip 10pt

    {\bf Constraints:} An {\em unconstrained} trading strategy starts and ends at pre-determined quantities but is otherwise free to hold any quantity. at any time. Constrained strategies on the other hand have limits on what quantities of stocks may be held that reflect so-called "real-world" considerations that include, among others, maximum size limits and short-selling constraints. In more detail here are several examples of such constraints:

    \begin{itemize}
        \item {\bf Restrictions on overbuying (see Section \ref{sec:constraint-overbuy})}: optimal position-building strategies can significantly {\em } relative to the target quantity. In either trader may be subject to restrictions on maximum position size which would lead to a constraint of the form, say, $a(t) \le 1 + \rho$ for all $t$; 

        \item {\bf Channel constraints (see Section \ref{sec:constraint-channel})}: a trader may wish to confine the trajectory of the trading strategy to a certain channel.

        \item {\bf End-strategy constraints (see Section \ref{sec:constraint-End-strategy}):} a trader may wish to have a certain fraction of the target quantity acquired by no later than a certain time.

        \item {\bf Short-selling constraints (see Section \ref{sec:constraint-short-selling}:} quite simply, sometimes a trader building a long position is prohibited from selling short altogether or beyond a certain size.

        \item {\bf No-sell constraints (see Section \ref{sec:constraint-no-sell}):} suppose that $A$ never wants to sell the stock, this translates to a constraint on the first-derivative, namely, $\dot a(t) \ge 0$ for all $t$.
    \end{itemize}

    \vskip 10pt
    
    {\bf Game theory and dynamics:} The two key game-theoretical concepts in this paper and \cite{arXiv:2409.03586v1} are best-response and equilibrium. A {\em best-response} strategy produces the most favorable outcome taking the strategies of other traders as given. This is a completely standard interpretation of the concept of best-response from game theory; similarly {\em equilibrium} is when two or more strategies are {\em simultaneously} the best-responses to each other\footnote{The literature on game theory is extensive and here we mention but a few including the theoretical starting point \cite{von2007theory}, Nash's seminal work introducing the concept that became known as Nash equilibrium in \cite{nash_1950}, \cite{nash_1951}, \cite{nash1950equilibrium} and \cite{nash1953two}. See also \cite{holt_roth_2004} for a broader perspective.}. As is noted in \cite{holt_roth_2004}:

    \begin{quote}
    \em 
    When the goal is prediction rather than prescription, a Nash equilibrium can also be interpreted as a potential stable point of a dynamic adjustment process in which individuals adjust their behavior to that of the other players in the game, searching for strategy choices that will give them better results.    
    \end{quote}

    In the case of position building in competition this is {\em exactly} what happens when we introduce computational methods for finding two-player equilibria. In Section \ref{sec:two-trader-equilibria-compute} we introduce an algorithm for computing two-trader equilibria that {\em in effect} replicates this dynamic adjustment process (see \ref{alg:two-trader-eq}). The algorithm allows one to understand the entire {\em path} leading to equilibrium and helps to explain certain features of equilibria such as in certain cases the cost of trading in equilibrium are higher than non-equilibrium, which can be seen as a direct consequence of the {\em non-cooperative} nature of the games.

    \subsection{Mathematical overview of trading strategies}

    In this paper we focus primarily on two-trader position building scenarios where there is a unit trading strategy $a$ and a $\lambda$-scaled strategy $\blam$, however the most general scenario is when there are $n$-traders trading strategies $a_i$, where $i$ represents trader $i$. In both cases strategies satisfy the boundary conditions of vanishing at $t=0$ and acquiring $\lambda_i$ units of stock at time $t=1$. In the two-trader setup $a=a_1$, $\blam=a_2$, $\lambda_1=1$, $\lambda_2=\lambda$.

    \vskip 10pt
    
    {\bf Market impact:} When $n$ traders are in competition it is assumed that their aggregate trading exerts the marginal influence on the price of the stock. A trader seeking to purchase a quantity of stock from time $t=0$ to $t=1$ is concerned with two forms of trading cost. One is the transitory cost associated with paying in excess of the prevailing market price at any given time, and the other is the cumulative cost recorded as the difference between the price prevailing at time $t=0$ and the price paid at the time of a given trade. For strategies $a_i$ we model the cost paid by trader $i$ at time $t$ as proportional to:

    \begin{equation}
        \left(\sum_j \dot{a}_j(t)\right) \dot{a_i}(t) + 
            \kappa \left(\sum_j a_j(t)\right) \dot{a_i}(t) 
    \end{equation}

    This cost says that relative to the price at time $0$ and executing every trade at the prevailing price at the time of trade depends on the {\em aggregate} quantity traded at time $t$ by all traders and the net quantity acquired from time $t=0$ to $t$. The parameter $\kappa$ is called the {\em market impact coefficient} and represents the relative contribution between the temporary and permanent impact terms. Many functions in the sequel are  parameterized by $\kappa$ and in those cases we will leave $\kappa$ as implicit to de-clutter the notation.

    \vskip 10pt

    {\bf Best-response and equilibrium strategies:} \cite{arXiv:2409.03586v1} studies finding the strategy $a$ that minimizes its total cost of trading in competition with a strategy $b_\lambda$ where the total cost of trading is given by:

    \begin{equation}
        \label{eq:cost-ab}
        \Costab=\intzo (\dot a + \lambda \dot \blam)\dot a + \kappa(a + \lambda \blam)\dot a \dt
    \end{equation}
    
    and therefore the best-response strategy is defined as the solution to:
    
    \begin{equation}
        \label{opt:best-response-unconstrained}
        \underset{\underset{a(0)=0, a(1)=1}{a:[0,1]\to \R}}{\minimize} \,\, \int_0^1 (\dot a + \lambda\dotblam)\dot a + \kappa (a + \lambda \blam) \dot a \dt
    \end{equation}

    In \cite{arXiv:2409.03586v1} a two-trader equilibrium was defined simply as a pair of strategies $(a, \blam)$ such that $a$ is the best-response to $\blam$ and $\blam$ is the best-response to $a$.

    \vskip 10pt

    {\bf Best-response and equilibrium strategies.} For $a$ trading in competition with a $b_\lambda$ the best-response strategy is the one that minimizes total-cost, which can be expressed as the integral of the loss function:

    \begin{equation}
        \label{eq:loss-function-ab}
        L_a(t; \blam) = (\dot a + \dotblam)\dot a + \kappa(a + \lambda \blam)\dot a
    \end{equation}

    and we have

    \begin{equation}
    \label{eq:loss-function-ba}
    L_{\blam}(t; a) = (\dot a + \lambda \dotblam) \lambda\dotblam + \kappa(a + \lambda \blam) \lambda\dotblam
    \end{equation}

    To find the cost-minimizing strategy we use the Euler-Lagrange equation:
    
    \begin{equation}
        \label{eq:euler-lagrange}
        \frac{\partial L_a}{\partial a} - \frac{\mathrm{d}}{\mathrm{d}t} \left( \frac{\partial L_a}{\partial \dot{a}} \right) = 0
    \end{equation}

    with the resultant PDE being:

    \begin{equation}
        \label{eq:optimal-a-pde-vs-ab}
        \ddot a = -\frac{\lambda}{2} ({\ddot{b}}_\lambda + \kappa \dotblam) 
    \end{equation}

    By the same token, if $\Costba = \intzo L_{\blam}(a, \kappa) \dt$ we may also use the Euler-Lagrange equation similarly to derive an expression for the minimal strategy $\blam$ trading in competition with $a$:

    \begin{equation}
        \label{eq:optimal-b-pde-vs-ba}
        {\ddot{b}}_\lambda = -\frac{1}{2\lambda} (\ddot a + \kappa \dot a)
    \end{equation}

    Finally in \cite{arXiv:2409.03586v1} it was shown that the system of system of equations \eqref{eq:optimal-a-pde-vs-ab} and \eqref{eq:optimal-b-pde-vs-ba} has a pair of solutions $a, \blam$ and this means that $a, \blam$ are {\em simultaneously} the best-responses to one another and is a two-trader equilibrium.

    {\bf Fourier approximations:} We use Fourier series to approximate trading strategies, which will be the computational engine used throughout this paper. Given a trading strategy $a$ we know that the function $a(t)-t$ vanishes at zero and one and therefore the Fourier series for  $a-t$ is a sine series. For unit strategies $a, b$ we write $a_\F, b_\F$ for their corresponding Fourier expansions and note the $n$-th coefficient may be computed via integration as follows:

    \begin{subequations}
        \begin{align}
        \label{eq:fourier-a-func}
        a_{\F}(t) &= t + \sum_1^N a_n \sin(n \pi t), \qquad a_n = \intzo (a(t)-t) \sin(n\pi t) \dt \\
        \label{eq:fourier-b-func}
        \blamF(t) &= t + \sum_1^N b_n \sin(n \pi t), \qquad b_n = \intzo (b(t)-t) \sin(n\pi t) \dt 
        \end{align}
    \end{subequations}

    Observe that $N$ is left implicit and from standard results in Fourier series we know that $a_\F\to a$ uniformly as $N\to \infty$ and the coefficients $a_n$ decay like $1/n^2$ for twice-differentiable functions; identical results hold for $b$. In making these observations we are able to identify the space of unit trading strategies with elements in $\R^N$:

    \begin{equation}
        a(t) \sim a_\F = t+\asin \mapsto (a_1, \dots, a_N) \in \R^N
    \end{equation}

    This identification will allow us to translate problems about the functions representing trading strategies to minimization problems over sets in $\R^N$. We use the next sections to do this.

    \subsection{Convex combinations of trading strategies and the cost function\protect\footnote{This section is new to Version 2. The prior version stated Algorithm \ref{alg:two-trader-eq} without any explanation of why it works.}}
    \label{sec:convex-combinations}

    Consider $0\le \gamma \le 1$ and two strategies $\blamA$ and $\blamB$ we can form the {\em convex combination}

    \begin{equation}
        \blamG = \gamma \blamA + (1\shortminus\gamma)\blamB
    \end{equation}

    and then for a strategy $a$ we have the associated cost functions $\Cost(a,\gamma \blamA)$, $\Cost(a, (1\shortminus \gamma)\blamB)$ and $\Cost(a, \blamG)$. It is immediate consequence of the linearity of integration and differentiation  and they are related as follows:

    \begin{equation}
        \label{eq:convex-combination-loss-fns}
        \Cost(a, \blamG) = \Cost(a, \gamma\blamA) + \Cost(a, (1\shortminus \gamma) \blamB) 
    \end{equation}

    The meaning of \cref{eq:convex-combination-loss-fns} is as follows. The left-hand side is the cost of $a$ trading in competition with $\blamG$, while the right-hand side is the sum of the cost of $a$ trading in competition with $\gamma\blamA$ and $a$ trading in competition with $(1\shortminus \gamma)\blamB$. Now suppose that $a^*$ is {\em optimal} versus $\blamG$, that is, that it minimizes $\Cost(a, \blamG)$.

    \section{Fourier series and trading strategies}

    Best-response strategies are the {\em results} of the minimizations in \ref{eq:cost-ab} involving $\Costab$, and similarly for $\Costba$. To translate this a minimization problem in $\R^N$ we proceed by replacing $a, b$ with $a_\F, b_\F$ in the loss functions $L_a$ and $L_b$ respectively and then the uniqueness of Fourier series combined with convergence and term-by-term integration results guarantees we can find a suitable approximation for the corresponding cost functions.

    \subsection{Approximating the cost function with Fourier series}
    \label{sec:fourier-cost-function}

    This section provides the computational backbone of this paper. We will show that the cost function is a quadratic function of the Fourier coefficients of the strategies $a, \blam$.

    \begin{prop}
    \label{prop:approx-cost-ab}
    The total cost function $\Costab = \int_0^1 (\dot a + \lambda\dot \blam)\dot a + \kappa (a + \lambda \blam) \dot a \dt$ may be approximated by the a function $\CostFab$:

    \begin{equation}
    \label{eq:fourier-cost-fn}
     \CostF(a, \blam) =
        \frac{1}{2}(2+\kappa)(1+\lambda) + \frac{\pi^2}{2} \sum_{n=1}^N  (a_n^2 + \lambda a_n b_n) n^2    
            +\frac{2\kappa\lambda}{\pi}   \sumodd \frac{b_n- a_n}{n} + 
    2\kappa \bigbs \sumnmodd  \frac{\lambda b_m a_n n m}{m^2 - n^2}             \end{equation}
    \end{prop}

    \vskip 10pt
    
    Similarly:

    \vskip 12pt

    \begin{prop}[Approximate cost function for $b$ with respect to $a$]
    \label{prop:approx-cost-ba}
    
    The approximate cost function $\CostFba$ is given by the formula:
    \begin{equation}
    \label{eq:fourier-cost-fn-ba-final}
        \CostFba = \lambda \left\{\frac{1}{2}(2+\kappa)(1+\lambda) \!+\!
            \frac{\pi^2}{2} \sum_{n=1}^{N}  \left(\lambda b_n^2 \!+\! a_n b_n \right)n^2 +
             \frac{2\kappa}{\pi} \sumodd \frac{a_n- b_n}{n}  \!+\! 
             2\kappa\bigbs \sumnmodd  \frac{ a_m b_n n m}{m^2 - n^2} \right\}
     \end{equation}
    
    We present the proof of this approximation in Section \ref{sec:proof-cost-fn-approx-ba}.
    
    \end{prop}

    \vskip 8pt

    \subsection{Properties of the approximate cost function}

    The approximate cost function $\CostFab$ in equation \eqref{eq:fourier-cost-fn} is quadratic in the Fourier coefficients of $a_\F$, and similarly $\CostFba$ is quadratic in the coefficients of $b_\F$. Also, an immediate consequence of \cref{eq:fourier-cost-fn} and \cref{eq:fourier-cost-fn-ba-final} is that the Hessian matrix of both are diagonal and positive definitive, which will imply that the quadratic programs we focus on when adding constraints will be convex. Both cost functions depend implicitly on the number of Fourier coefficients used to approximate $a_\F$ and $b_\F$ and we can show that as $N\to \infty$ the approximate cost functions converge to the actual cost functions. We state this now and sketch the proof.

    \vskip 10pt

    \begin{prop}[The approximate cost-function converges to the cost-function]
    Let $a, a_\F$ be as above, noting that $a_\F$ depends implicitly on the number of Fourier coefficients $N$. Then

    \begin{equation}
        \CostFab \to \Costab, \quad N\to \infty
    \end{equation}

    The analogous statement holds for $\CostFba$.        
    \end{prop}

    \vskip 10pt

    {\bf Sketch of proof:} To see this we use two key facts about Fourier series. The Fourier coefficients of a twice-differentiable function are unique and the integral of a partial Fourier series is the sum of the integrals of the partial sums and that sum converges to the integral of the function.  

    \subsection{Some useful facts about Fourier series of trading strategies\protect\footnote{This section is new to Version 2.}}
    \label{sec:useful-facts-Fourier}

    There are a number of basic properties of the Fourier series of trading strategies that we note here.

    \vskip 10pt

    {\bf Convex combinations of trading strategies.} Given unit strategies $a_1, a_2$ and $0\le \gamma \le 1$, $a_{12}= \gamma a_1 + (1\shortminus\gamma) a_2$ is a {\em convex combination} of the strategies. Since $a_{12}(0)=0$ and $a_{12}(1)=1$, it has a Fourier series: 

    \begin{equation}
    \label{eq:convex-fourier-1}
        a_{12,\F} = t + \sum_{n=1}^\infty a_{12, n} \sin(n\pi t) 
    \end{equation}

    We can also form the Fourier series of the function $\gamma a_1 + (1\shortminus\gamma) a_2$:

    \begin{equation}
    \label{eq:convex-fourier-2}
        \gamma a_1 + (1\shortminus\gamma) a_2 = t + \sum_{n=1}^\infty \left(\gamma a_{1, n} + (1-\gamma) a_{2, n}\right) \sin(n\pi t) 
    \end{equation}

    and it is immediately clear that the series in \cref{eq:convex-fourier-2} the expression in \cref{eq:convex-fourier-1} and \cref{eq:convex-fourier-2} must be equal which immediately implies:
    
    \begin{equation}
        \gamma a_{1, n} + (1\shortminus\gamma) a_{2, n} = a_{12, n}, \qquad \text{for all $n$}
    \end{equation}
    
    We will use this fact in Section \ref{sec:two-trader-equilibrium}. 
    
\section{Constraints}

    In this section we describe some of the types of constraints we can use impose on the strategies we calculate and explain the strategy we employ to make these calculations. 

    \subsection{How to constrain best-response and equilibrium strategies}

    We outline the steps steps to achieve this as follows. Since each of the two trading strategies $a(t), b(t)$ satisfy the boundary conditions that they vanish at 0 and are 1 at time $t=1$, they may be approximated by sine series in $N$ terms
    
    \begin{equation}
        a(t) \sim a_\F = t+\asin, \qquad b \sim b_\F =  t+\bsin
    \end{equation}
    
    where $a_n = 2 \intzo (a(t) -t) \sin(n \pi t)$, $b_n=2\intzo (b(t)-t) \sin(n\pi t) \dt$, respectively. With this in hand we may identify each strategy with an element of $\R^N$:

    \begin{equation}
        a(t) \mapsto (a_1,\dots, a_N)\in \R^N \qquad \blam \mapsto (b_1,\dots, b_N)\in \R^N
    \end{equation}

    and therefore we may approximate the loss function $L(t)$ as follows:

    \begin{equation}
        L(t) = (\dot a + \dotblamF) + \kappa(a + \lambda \blam)\dot a
        \xmapsto{\hspace{1cm}} L_\F(t) = (\dot a_\F + \lambda \dotblamF) + \kappa(a_\F + \lambda \blamF)\dot a_\F
    \end{equation}

    By virtue of uniqueness of Fourier series for twice-differentiable functions, $L_\F$ is the Fourier series in $N$ terms for $L(t)$. Finally by the fact we can perform term-by-term differentiation of Fourier series we can identify the cost function $\Costab$ with an approximation:

    \begin{equation}
        \Costab = \intzo L(t) \dt
        \xmapsto{\hspace{1cm}} \CostF(a, \blam) = \intzo L_\F(t) \dt 
    \end{equation}

    Since the loss function itself is at most degree two, it is clear that $\CostFab$ is at most a degree-two polynomial in the Fourier coefficients of $a$ and $b$. It is the content of Section \ref{sec:fourier-cost-function} to derive this function and give an explicit formula for $\CostFab$ as a quadratic function in the coefficients $a_1\dots a_N$ and $b_1\dots b_N$. From this it becomes immediately obvious that $\CostFab$ is also positive-definite and it is straightforward to show it converges uniformly to $\Costab$ as the number of terms $N\to \infty$. It is with this identification that we may translate problems relating to minimizing $\Costab$ to quadratic programs $\CostFab$.
    
    \begin{center}
        \begin{tabular}{p{5.5cm}p{2.5cm}p{5.5cm}}
    	{\begin{equation*}
    			\begin{array}{lll}
    				\text{(EL)}
    				&\minimize & \Cost(a, \blam)\\
                        &\text{subject to}&\textit{No constraints}
    			\end{array}
    	\end{equation*}}
            &
            $$
            \begin{array}{c}
        	\xleftrightarrow{\hspace{2cm}}
            \end{array}
            $$
            &
    	{ \begin{equation*}
    			\label{QP}
    			\begin{array}{lll}
    				\text{(QP)}&\minimize& \CostF(a, \blam)\\
    				&\text{subject to}& \textit{Constraints}
    			\end{array}
    	\end{equation*}}
        \end{tabular}
    \end{center}

    where the minimizations of $a$ and $\blam$ respectively. With the basic approach in-hand we next demonstrate how to how to introduce constraints.

    \vskip 10 pt

    {\bf Approximating an interval with a finite set of points:} Our aim below is to constrain a function $f$ on an interval $I_{ab}=[a, b]$ {\em well enough} that the size of the set of points where a constraint such as $f(t) < g(t)$ for all $t\in I_{ab}$ fails to hold is sufficiently small. Thus we want a set $\T = \{t_1, \dots, t_K\}\subset [a, b]$ such that 
    
    $$
    \mu \left(\{t\in I_{ab} \,|\, f(t) > g(t)\,\}\right) < \epsilon
    $$
    
    where $\mu$ is some measure on the real numbers, e.g., Lebesgue measure. We omit the details here, and simply note that for relatively "well-behaved" functions and constraints we can construct sufficient sets for our purposes. In what follows we will simply refer to an {\em approximating set of times} with respect to a function (or functions).

    {\bf Path constraints on trading strategies.} Given a function $c:[0, 1] \to \R$, a strategy $a$ is said to satisfy the {\em path constraint} defined by $c$ if $a(t) \le c(t)$ for all $t\in [0, 1]$. We call $c$ the {\em constraint function}. For an approximating set of times $\T = \{t_1, \dots, t_K\}\subset [0, 1]$ with respect to $M$ constraint functions $c_1, \dots, c_M$ we consider the following constrained optimization:
    
    \vskip 10pt

    \begin{equation}
      \begin{array}{ll}
        \underset{(a_1\cdots a_n)\in \R^N}{\minimize}   & \quad \CostF(a, \blam) \\[2.0ex]
        \text{subject to} & \quad t + \asin \le c_i(t), i=1, \dots, M \quad \text{for all $t \in \T$ and $c_1, \dots c_M$}
      \end{array}
      \label{opt:cost-constrained}
    \end{equation}

    This is a quadratic program in $N$ variables with $K\times M$ {\em linear} constraints. In the rest of this section we demonstrate how to implement a variety of useful constraints using \eqref{opt:cost-constrained} but first introduce {\em partial path constraints}.

    \vskip 10pt

    {\bf Partial path constraints:} We can make a simple modification of path constraints by choosing an approximating set of times $\T^* = \{t_1, \dots, t_k\}\subset [t^*, 1]$, where $t^*>0$ with respect to some constraint functions $c_1, c_2, \dots$ to mean that the constraints only apply to some of the times. We now discuss a selection of path and partial path constraints that arise in practice.

    \vskip 10pt

    \subsection{Overbuying constraint}
    \label{sec:constraint-overbuy}

\begin{figure}[h!]
        \centering
        \includegraphics[width=0.5\linewidth]{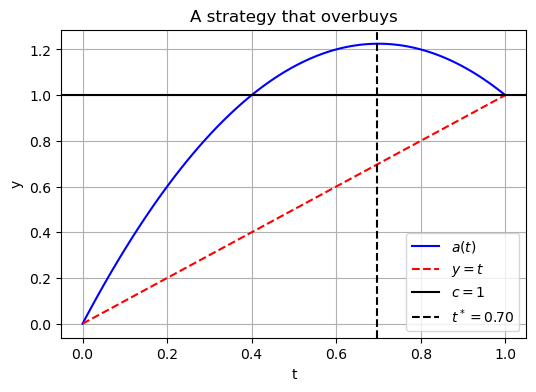}
        \caption{An example of a over-buying strategy (blue line) exceeds the target quantity of one and therefore violates the overbuy constraint.}
        \label{fig:over-buying-strategy}
    \end{figure}

    An  constraint is a path constraint $c(t) = 1+\rho$ for $\rho>0$ so that $a(t) \le 1 + \rho$ and thus limits the extent to which a best-response strategy can "overbuy" relative to its target quantity of $1$. Figure \ref{fig:over-buying-strategy} depicts this in general and Figure \ref{fig:best-response-risk_neutrall-overbuy} shows an example the best-response to a risk-neutral adversary with overbuy constraints.

    \subsection{Channel constraints}
    \label{sec:constraint-channel}

\begin{figure}[h!]
        \centering
        \includegraphics[width=0.5\linewidth]{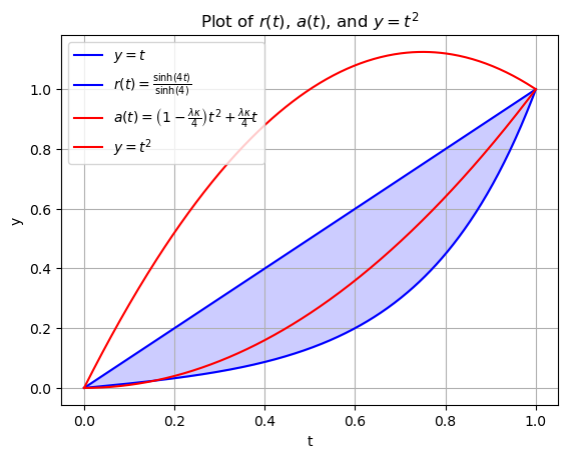}
        \caption{The trajectory channel bounded by $\sinh(4t)/\sinh(4)$ below and $t$ above. We may seek a constrained optimization that restricts to strategies that are within this channel and see that the risk averse strategy $y=t^2$ is always within the channel while the eager strategy is never within the channel.}
        \label{fig:trajectory-channel-example}
    \end{figure}

    A trajectory channel constraint seeks to keep the trading trajectory within some bounds defined by a pair of functions $L, U$ such that $L(t) \le U(t)$. A strategy satisfies the channel constraint if $L(t) \le a(t) \le U(t)$ for all $t\in [0, 1]$, which is clearly an example of a path constraint. For example, if a traders wants the best-best response to some strategy $\blam $ but wants to be no more risk averse than $\sinh(4t)/\sinh(4)$ and take no more risk than a risk-neutral strategy then this is defined by $L(t) = \sinh(4t)/\sinh(4)$ and $U(t)=t$. Figure \ref{fig:trajectory-channel-example} gives a illustration of this.

    \subsection{End-strategy constraints}
    \label{sec:constraint-End-strategy}

\begin{figure}[h!]
        \centering
        \includegraphics[width=0.5\linewidth]{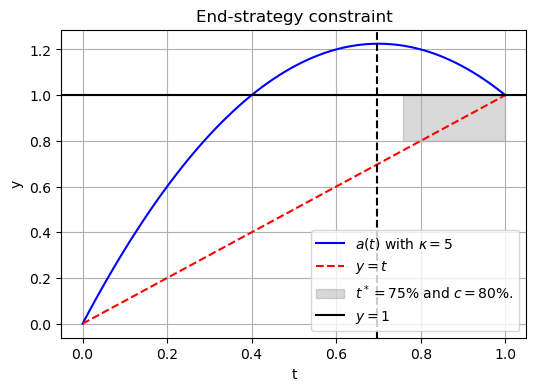}
        \caption{A visual depiction of a end-strategy constraints with $t^*=75\%$ and $c=80\%$. The constraint requires that at time $t=0.75$ the trading strategy is at least 80\% completed and does not exceed 100\% of the target quantity. The eager strategy (blue line) nor the risk-neutral strategy (red-line) satisfy the constraint. See figure \ref{fig:best-response-eager-End-strategy_sigma3} for examples of best-response strategies with end-strategy constraints.}
        \label{fig:End-strategy-constraints}
    \end{figure}

    End-strategy constraints are examples of {\em partial path constraints}, that is, constraints that only apply to a subset of times, in this case in the interval $[t^*, 1]$ for time $t^*>0$. In this case each path constraint  pertains to a subset of times $\T^*=\{t_1, \dots, t_K\}\subset [t^*, 1]$. In practice, end-strategy constraints require a strategy to exhibit some behavior from time $t^*$ to time $t=1$, the start of the "end-strategy". We illustrate the most basic end-strategy constraint, that for some $c<1$, $c \ge a(t) \le 1$ for $t\in [t^*, 1]$. We depict this constraint in Figure \ref{fig:End-strategy-constraints}, and an illustration is in Figure \ref{fig:End-strategy-constraints}

    \subsection{Short-selling constraints}
    \label{sec:constraint-short-selling}

    Short-selling constraints either prohibit short-selling altogether (strict short-selling constraint) or limit it in some way. For both versions of short-selling constraints the problem simple. Choose time points $t_1,\dots t_K$ and $c\le 0$ and solve: 

    \begin{equation}
       \begin{array}{ll}
         \minimize   & \quad \CostFab\\[1.25ex]
         \text{subject to} & \quad t + \sum_{n=1}^N a_n \sin(n \pi t) \ge c,\, t\in \T 
      \end{array}
    \end{equation}

    For strict short-selling constraints set $c=0$ and non-strict constraints $c < 0$. This simply places a floor below which holdings cannot go. Note that $c < 0$ means since $a(t)$ is trading a unit strategy then $|c|$ expresses the short-selling constraint as a fraction of the target size.

    \subsection{No-sell constraints}
    \label{sec:constraint-no-sell}

    In \cite{arXiv:2409.03586v1} it was noted that a strategy shape that arises in the study of position building with competition is the {\em bucket strategy} depicted in Figure \ref{fig:over-buying-strategy}. Suppose that the trader wishes to constrain solutions to optimal best-response strategies to ones that do not trade bucket strategies.

    There are two approaches to this. The first is simply to constrain the first derivative of the best response strategy to be non-negative, that is, to insure that $\dot a(t) \ge 0$ for all $t\in [0,1]$. We can achieve this using the Fourier approximation of $\dot a$:

    \begin{equation}
        \dot{a}_\F(t) = 1 + \sum_{n=1}^N a_n\, n \pi \cos(n \pi t) 
    \end{equation}

    and for a set of times $t_1\cdots t_K$ require $\dot{a}_\F(t_k) \ge 0$ for each $k$. This can then be solved using

    \begin{equation}
    \label{opt:cost-no-sell}
        \begin{array}{ll}
         \minimize   & \quad \CostFab \\[1.25ex]
         \text{subject to} & \quad t+\acos \ge 0,\, t\in \T
        \end{array}
    \end{equation}

    \vskip 12pt

    No-sell constraints by definition constrain a strategy from selling at {\em any point} time along the trajectory, and therefore, will automatically prevent . However, no-sell constraints are {\em much stronger} overbuy constraints because they prohibit selling when the strategy is holding less than the target quantity as well as when it holds more. By contrast, overbuy constraints prevent selling as a {\em consequence} of the target quantity. If a strategy holds more than the target quantity of $1$ at any time, then selling is required to ultimately satisfy the boundary condition.

\section{Numerical examples of strategies with constraints}

    In the sections that follow we provide numerical examples of best-response and equilibrium strategies with constraints. We run through a collection of strategies that a single adversary might trade.\footnote{The code for all of these examples can be found online at \url{https://github.com/vragulin/Optimal_Exec_In_Comp}, commit 7bc0c74.} To explain this start by noting that {\em every} adversary depends on a scaling parameter $\lambda$, those strategies that take into account the market impact coefficient $\kappa$ will depend on $\kappa$ and those arise when, for instance, the strategy is itself a best-response strategy. Finally many strategies are themselves members of parametric families, for example risk-averse strategies depend on a risk aversion parameter, $\sigma$ (see Section \ref{sec:best-response-risk-averse}).  

    \subsection{A taxonomy of trading strategies}
    \label{sec:taxonomy-strategies}

    To organize our thinking we start with a hierarchy of position-building trading strategies, in scenarios where there is a unit strategy $a$ in competition with a $\lambda$-scaled strategy $\blam$.

    \begin{enumerate}
        \item {\bf\em Passive:} Strategies that are not focused on competition and therefore they are parameterized only by their own parameters and do not require knowledge of the market impact parameter $\kappa$. {\em We include passive strategies in this taxonomy as the foundation of best-response strategies.}
        \begin{itemize}
            \item {\bf Risk-averse:} These are strategies that prioritize delaying purchases in order to minimize risk at the expense of higher market impact costs. Optimal strategies of this type have the form $\blam(t)=\sinh(\sigma t)/\sinh(\sigma)$, where $\sigma>0$ is a risk-aversion parameter.
            
            \item {\bf Risk-neutral:} These are strategies that balance spreading out trades and the expense of higher risk exposure. The optimal strategy in this category is $b(t)=t$; and
            
            \item {\bf Eager:} These are strategies that prioritize buying stock quickly in order to {\em trade ahead} of an adversary. Strategies in this category have the form $b(t) = \tfrac{e^{-\sigma t} - 1}{e^{-\sigma} - 1}$ where $\sigma$ is an eagerness parameter.
        \end{itemize}
        
        \item {\bf\em Best-response:} are optimal response strategies when trading in competition {\em with} some other strategy. The operative question when investigating a best-response strategy is {\em what is it the best response to?} These strategies require knowledge of both what the strategy that is being responded to is, and the market impact coefficient $\kappa$. Therefore, as a starting place each passive strategy has a best-response strategy.
        
        \begin{itemize}
            \item {\bf Risk-averse:} Best-response strategies to risk-averse strategies and are discussed in Section \ref{sec:best-response-risk-averse}.
            
            \item {\bf Risk-neutral:} Best-response strategies to risk-neutral strategies and are discussed in Section \ref{sec:best-response-risk-neutral}.
            
            \item {\bf Eager:} Best-response strategies to eager strategies were discussed in detail in \cite{arXiv:2409.03586v1} discussed again in this paper in \ref{sec:best-response-eager}. An noteworthy feature of these is that when the level of eagerness is high, they can respond by {\em selling short}; some practitioners find this counter-intuitive and do not want to sell short, so in Section \ref{sec:best-response-eager} we illustrate the no-sell constraint. 
            
        \end{itemize}
        \item {\bf\em Equilibrium:} These are strategies arise from stable sets of strategies that are pairwise best-responses to one another. In all cases, we focus on a unit strategy that is the equilibrium best-response to one or more other strategies. 
            \begin{itemize}
                \item {\bf Two-trader:} As described in the introduction an equilibrium is defined by a pair of strategies that are simultaneously the best responses to one another;

                \item {\bf $n$-trader symmetric:} This concerns the situation where $n$ traders all wish to purchase the same quantity of stock.
            \end{itemize}
    \end{enumerate}

    \subsection{Best-response to a risk-neutral adversary}
    \label{sec:best-response-risk-neutral}

    This type of strategy is the best-response to a trader $B$ who trades $\lambda$-scaled ($\lambda \ge 1$) risk-neutral strategy $\blam=t$. The corresponding best-response strategy is given by

    \begin{equation}
        \label{eq:best-response-risk-neutral}
        a(t) = \left(1 + \frac{\lambda \kappa}{4}\right) t - \frac{\lambda \kappa}{4} t^2     
    \end{equation}

\begin{figure}[h!]
        \centering
        \includegraphics[width=0.75\textwidth, height=0.5\textwidth]{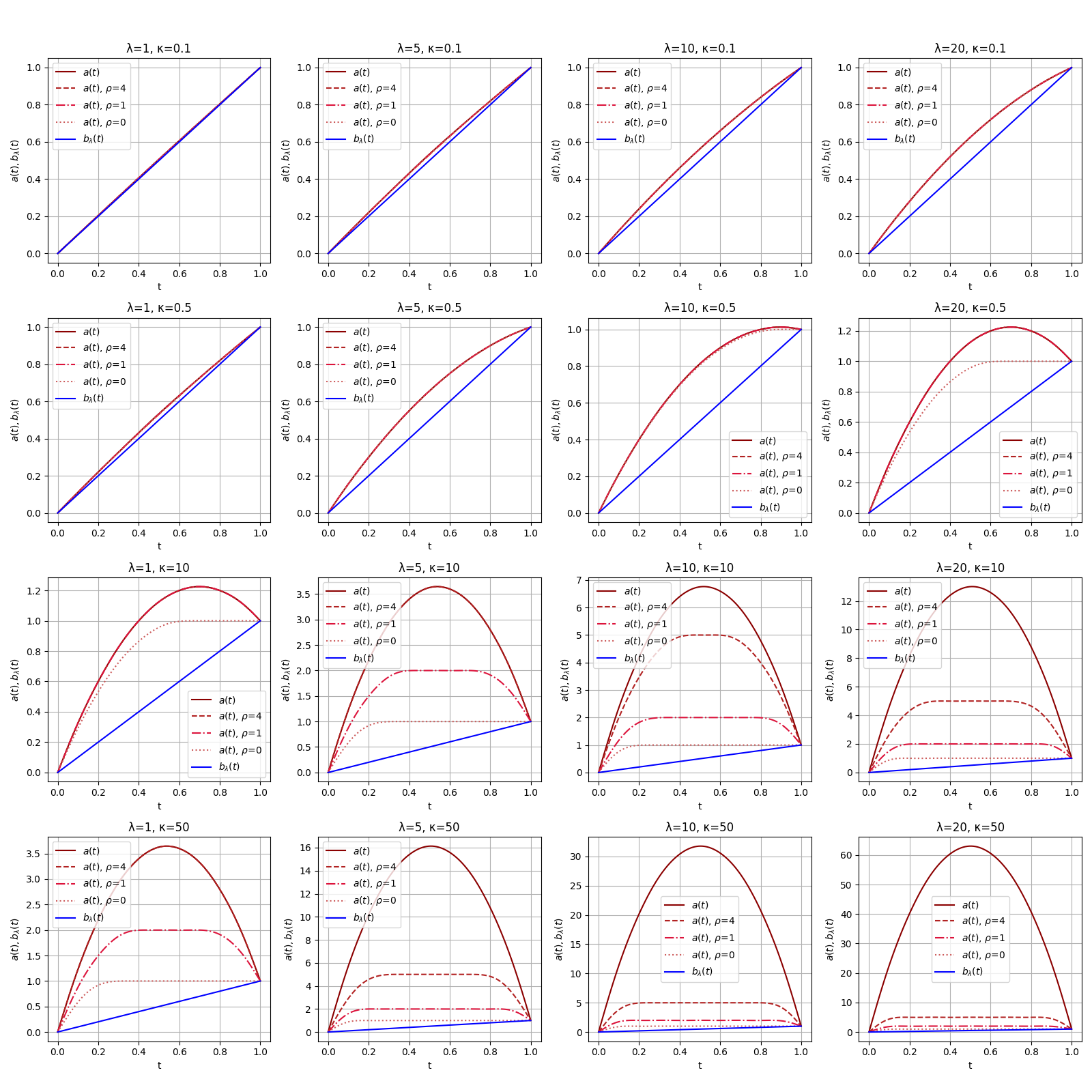}
        \caption{Best-response strategies $a$ to a risk-neutral adversary trading $b_\lambda$ with an overbuying constraint. The plots depict two different overbuying buying constraints (500\% and 200\% of the target quantity) in dashed red and the unconstrained in solid red.}
        \label{fig:best-response-risk_neutrall-overbuy}
    \end{figure}

    Figure \ref{fig:best-response-risk_neutrall-overbuy} shows plots of the best-response to a risk-neutral competitor with some constraints on over-buying.

    \subsection{Best response to a risk-averse adversary}
    \label{sec:best-response-risk-averse}

\begin{figure}[h!]
        \centering
        \includegraphics[width=0.75\textwidth, height=0.5\textwidth]{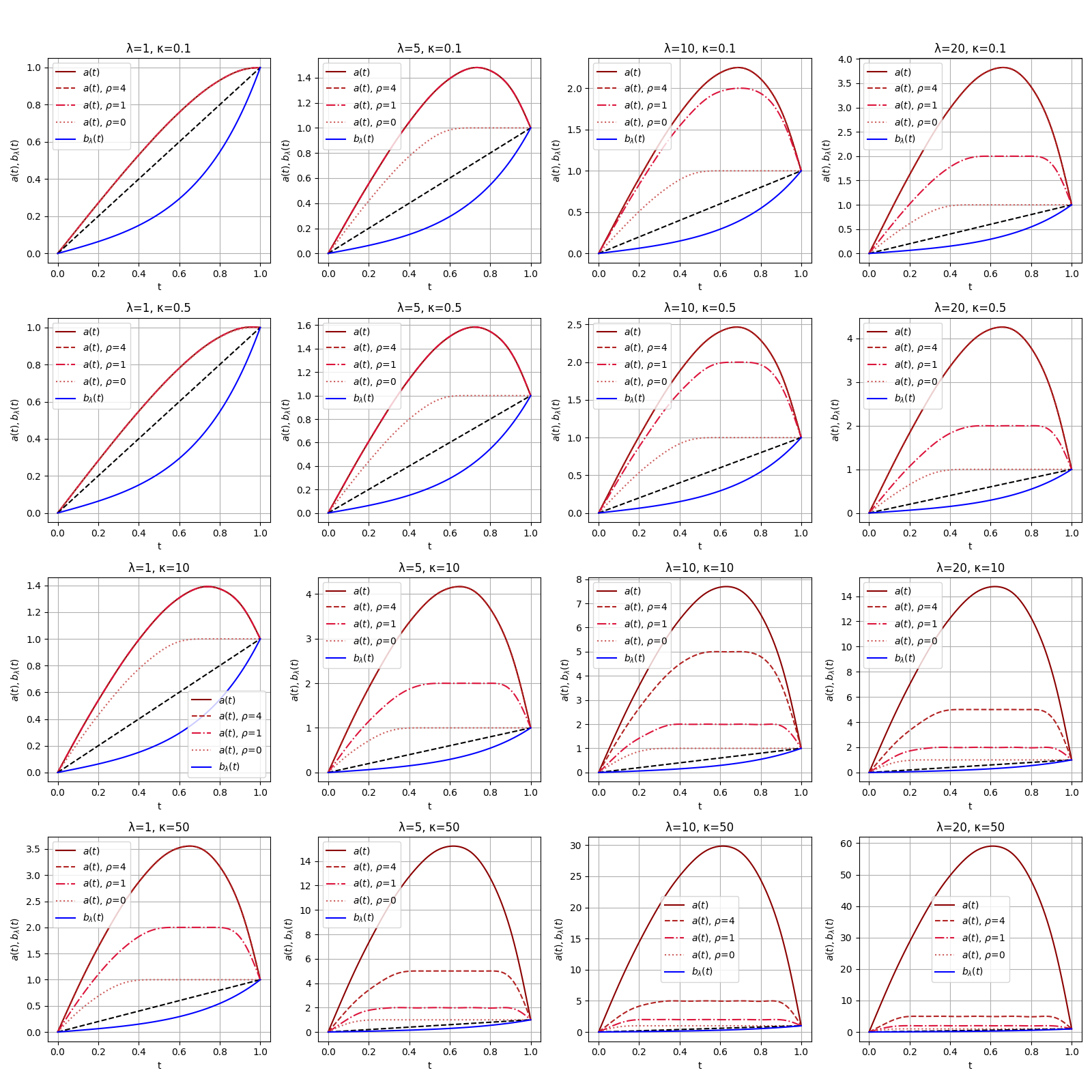}
        \caption{Best-response strategies $a$ to a risk-averse competitor trading $\blam$ with risk-aversion parameter $\sigma$ as in \cref{eq:best-response-strategy-ra-neat} for $\sigma=3$. The results are shown for both unconstrained (solid red) and with the  constrained for various values of $\rho$.}
        \label{fig:best-response-risk_averse_sigma3}
    \end{figure}

    In this section we illustrate the best-response to a risk-averse trader trading an optimal {\em passive} strategy parameterized by $\sigma$. Write $\blam(t)$ the $\lambda$-scaled strategy of risk-aversion $\sigma$. In \cite{arXiv:2409.03586v1} it was shown that best response to a $\lambda$-scaled risk-averse strategy is given by a strategy $a(t)$ (that also depends on $\lambda$ and $\kappa$, which we leave implicit). 

    \begin{equation}
    \label{eq:best-response-strategy-ra-neat}
        a(t) = \dfrac{\lambda}{2}(q_\sigma(0)- q_\sigma(t)) + \big(1+\dfrac{\lambda}{2} (q_\sigma(1) - q_\sigma(0) )\big)\, t 
    \end{equation}

    where $q_{\sigma}(t)$ is an auxiliary function depending on $\sigma$ and $\kappa$: 

    \begin{equation}
        \label{eq:qt-fn}
        q_{\sigma}(t) = \frac{\sinh(\sigma \, t)}{\sinh(\sigma)} + \frac{\kappa}{\sigma}\cdot \frac{\cosh(\sigma  \, t)}{\sinh(\sigma )}
    \end{equation}
    
    where $q_{\sigma, 1} = q_\sigma(1)$ and $q_{\sigma, 2}=q_\sigma(0)$. Figure \ref{fig:best-response-risk_averse_sigma3} shows best-response strategies $a(t; \sigkapplam)$ from \cref{eq:best-response-strategy-ra-neat}.

    \subsection{Best response to an eager adversary}
    \label{sec:best-response-eager}
    
    Finally we examine the case of the best-response strategy to an eager competitor. For this trader we use the $\lambda$-scaled strategy as follows:

    \begin{equation}
        \blam(t) = \frac{e^{-\sigma t} - 1}{e^{-\sigma} - 1}, \qquad\text{eager $\lambda$-scaled, $\sigma$ eagerness}        
    \end{equation}
    
    as above we solve for the best-response strategy $a(t)$ with boundary conditions $a(0)=0, a(1)=1$ to obtain\footnote{In the version of \cite{arXiv:2409.03586v1} submitted September 5, 2024 there was an error in the expression for $\blam(t; \kappa, \sigma)$, the parameters $\sigma$ and $\lambda$ were taken as a single parameter.}:

    \begin{equation}
        \label{eq:best-response-eager}
        \blam(t) = \frac{e^{\sigma  (-t)} \left(\lambda e^{\sigma } (\sigma -\kappa )-t e^{\sigma  t}
            ((\lambda +2) \sigma -\kappa  \lambda )+e^{\sigma+\sigma  t} (-\lambda  \sigma +\kappa  (\lambda
            -\lambda  t)+(\lambda +2) \sigma  t)\right)}{2\left(e^{\sigma }-1\right) \sigma }
    \end{equation}

\begin{figure}[h!]
    \centering
        \includegraphics[width=0.75\textwidth, height=0.5\textwidth]{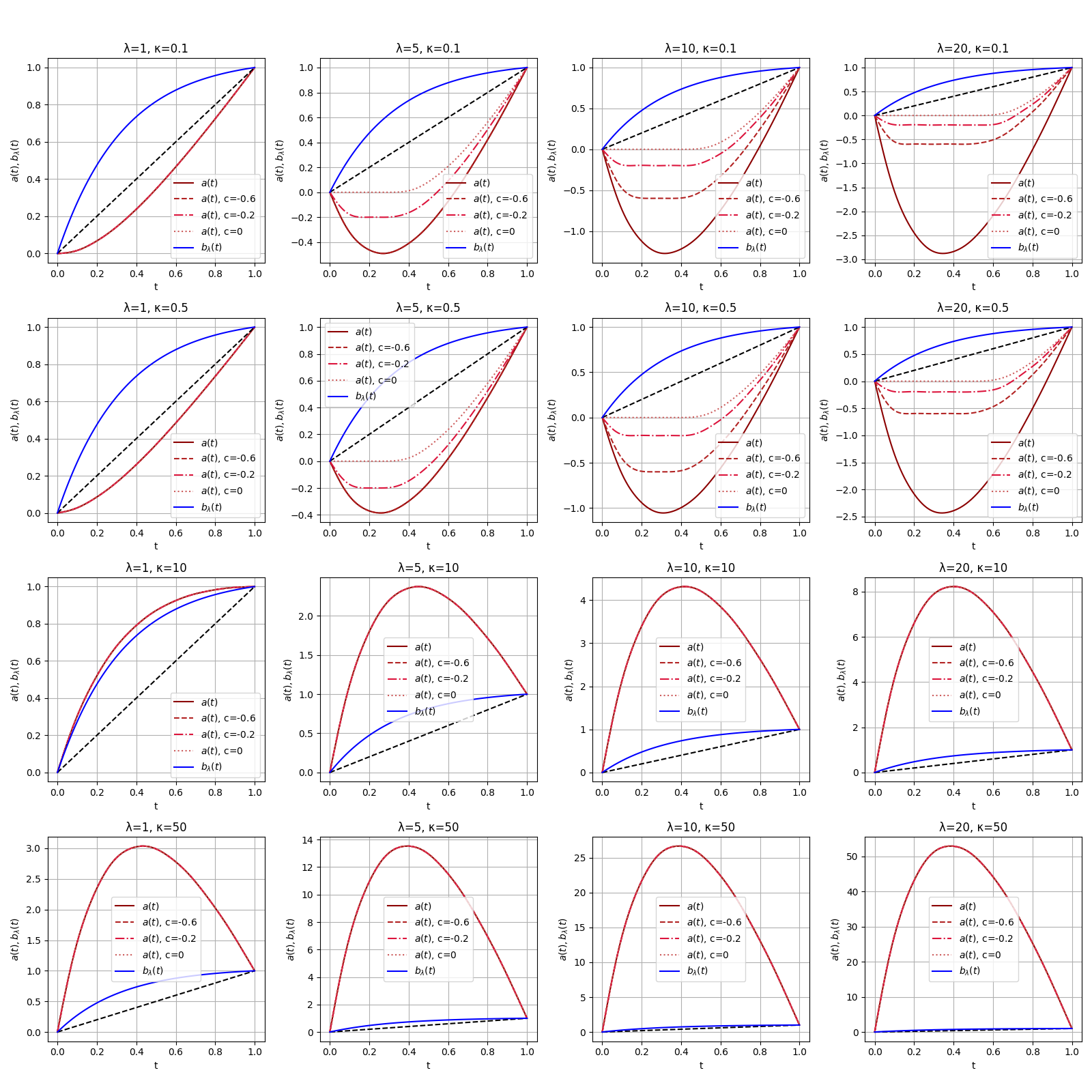}
        \caption{Best-response strategies to an eager competitor $\blam(t)$ (in blue) with eagerness parameter $\sigma$=3 as in \cref{eq:best-response-eager} for all plots. The best-response strategies are in solid and dashed red, where the solid red lines are unconstrained and the dashed red are various constraints on Short-Selling, see Section \ref{sec:constraint-short-selling}. In these plots the trader is not allowed to establish a position below $c$ at any time, i.e. $a(t) \geq c,  \,\text{for all } t \in [0,1]$.} 
        \label{fig:best-response-eager_sigma3_no_sell}
    \end{figure}

    To illustrate best-response to an eager strategy, in Figure \ref{fig:best-response-eager_sigma3_no_sell} we plot $\blam(t; \kappa, \sigma)$ for $\sigma=3$ of \cref{eq:best-response-eager} but as opposed to above we plot with the {\em no-sell constraint} (see Section \ref{sec:constraint-no-sell} due to the high propensity to sell short in these best-responses.

\begin{figure}[h!]
    \centering
        \includegraphics[width=0.75\textwidth, height=0.5\textwidth]{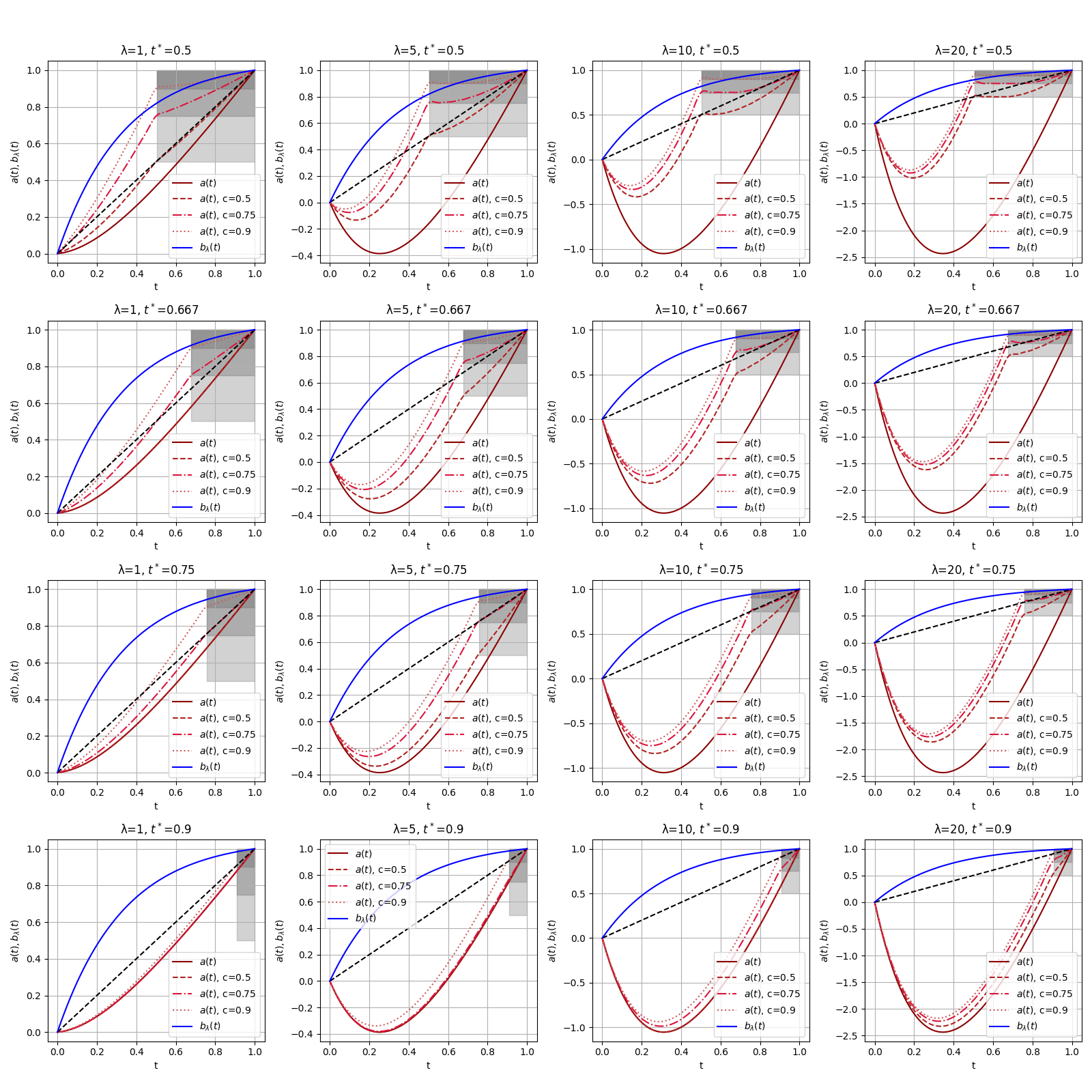}
        \caption{Best-response strategies to a passive eager competitor trading $\blam$ with eagerness $\sigma$ (see \cref{eq:best-response-eager}) for $\sigma=3$ and $\kappa=0.5$ computed with a variety of end-strategy constraints depicted by the grey shaded boxes (see Section \ref{sec:constraint-End-strategy}).}
        \label{fig:best-response-eager-End-strategy_sigma3}
    \end{figure}

    Figure \ref{fig:best-response-eager-End-strategy_sigma3} shows the best response for the same traders, but with a {\em end-strategy constraint} (see Section \ref{sec:constraint-End-strategy}).

    \subsection{Two-trader equilibrium closed-form solutions}
    \label{sec:two-trader-equilibrium}

    In \cite{arXiv:2409.03586v1} two-trader equilibrium strategies were derived as strategies $a$, $b_\lambda$ that are simultaneously the best-responses to one another given the market impact coefficient $\kappa$. The results are the equilibrium strategies $\aeq$, $\beq$ as follows:

    \begin{subequations}
    \begin{align}
        \aeq(t) &= -\frac{\left(1 - e^{-\tfrac{\kappa  t}{3}}\right)\!\!\left(-e^{\tfrac{\kappa}{3}} \left(e^{\tfrac{\kappa}{3}}+e^{2 \tfrac{\kappa}{3}}+1\right) (\lambda +1)+(\lambda -1) e^{\frac{\kappa  t}{3}}+(\lambda -1) e^{\frac{2 \kappa  t}{3}}+(\lambda -1) e^{\kappa  t}\right)}{2 \left(e^{\kappa }-1\right)}  \label{eq:two-trader-eq-a} \\
        \beq(t) &= \frac{\left(1 - e^{-\tfrac{\kappa  t}{3}}\right)\!\! \left(e^{\tfrac{\kappa}{3}} \left(e^{\tfrac{\kappa}{3}}+e^{2 \kappa/3}+1\right) (\lambda +1)+(\lambda -1) e^{\frac{\kappa  t}{3}}+(\lambda -1) e^{\frac{2 \kappa  t}{3}}+(\lambda -1) e^{\kappa t}\right)}{2 \left(e^{\kappa }-1\right) \lambda }  \label{eq:two-trader-eq-b}
    \end{align}
    \end{subequations}

\begin{figure}[h!]
    \centering
        \includegraphics[width=0.75\textwidth, height=0.5\textwidth]{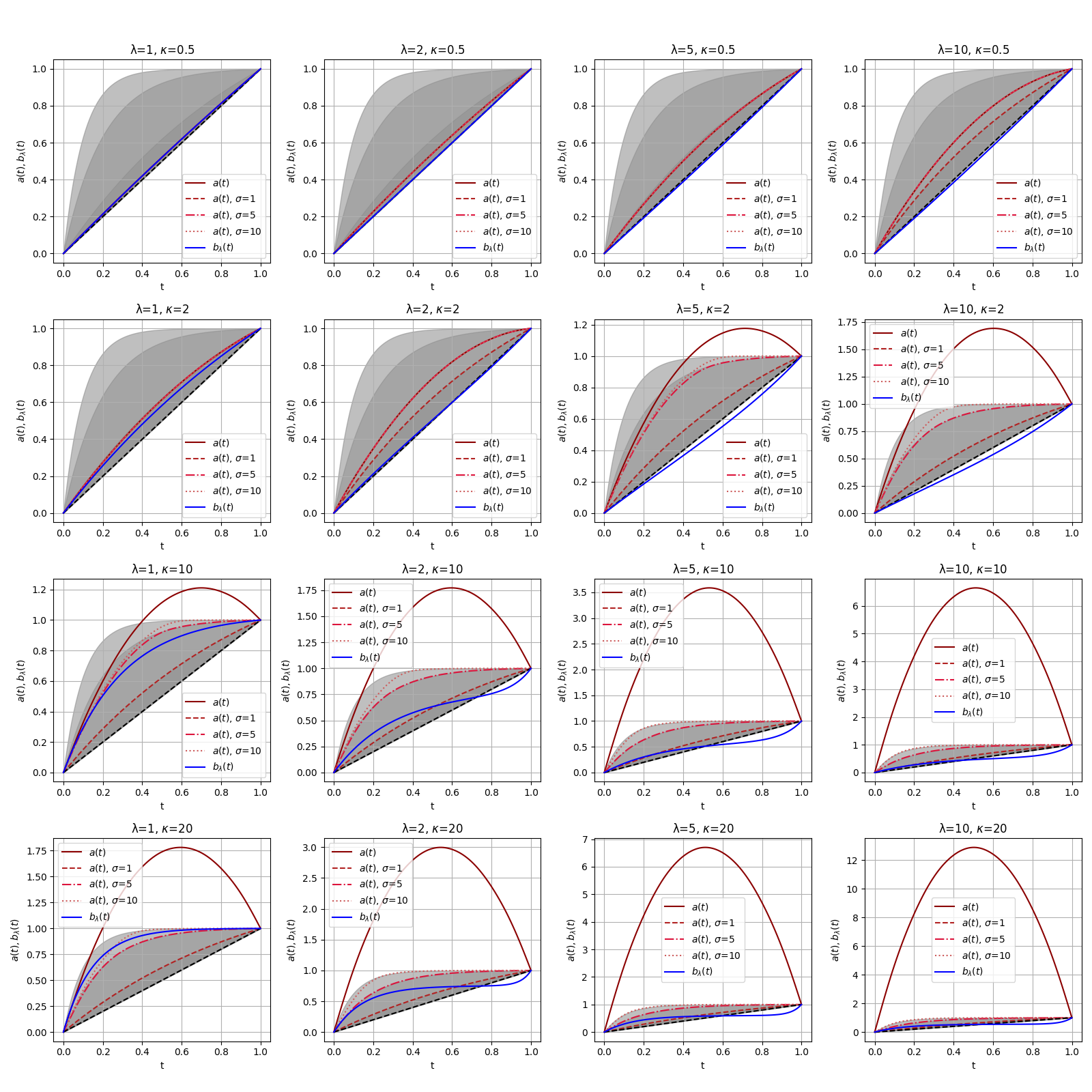}
        \caption{Best-response strategies to the two-trader equilibrium strategy $\blam$ for various values of $\lambda$; the best-response $a$ is constrained by the channel constraint set at various levels as depicted by the shaded grey areas (see Section \ref{sec:constraint-channel}). The channel constraints require that $a$ is always between the risk-neutral lower bound and the specified eager benchmark upper bound (\ref{sec:best-response-eager}).}
        \label{fig:best-response-to_tte_channel}
    \end{figure}

    If $A$ thinks that $B$ will trade the two-trader equilibrium strategy \cref{eq:two-trader-eq-b} for a given $\kappa$ and $\lambda$ then $A$'s best-response is to trade $\aeq(t)$. If, however, $A$ requires constraints on its strategy then while $A$ can add a constraint and solve for the best-response to $\blam$, this will {\em not} be an equilibrium strategy, because $B$'s strategy is not the best response to $A$'s constrained best-response to $B$. Nevertheless, Figure \ref{fig:best-response-to_tte_channel} depicts these best responses.

    \section{Computing two-trader equilibrium via iterative quadratic programming}
    \label{sec:two-trader-equilibria-compute}

    In \cite{arXiv:2409.03586v1} it was shown that two-trader equilibrium strategies exist by solving a system with partial differential equations with the Euler-Lagrange equation pertaining to each trader. As a very brief review, the paper derived equations defining best-response strategies for $A$ and $B$ position-building in competition where $B$ trades a $\lambda$-scaled strategy. The loss functions for $A$ and $B$ were as follows:
    
    \begin{align}
        L(a, \blam) &= (\dot a + \dotblamF) \cdot \dot a + \kappa (a + \lambda \blam) \cdot \dot a \label{eq-loss-fn-a}\\
        L(\blam, a) &= (\dot a + \dotblamF) \cdot \lambda\dotblam + \kappa (a + \lambda \blam) \cdot \lambda  \dot b \label{eq-loss-fn-b}
    \end{align}

    where $L(\blam, a)$ is the loss function $(\dot a + \dotblamF) \dotblamF + \kappa(a + \lambda \blam) \dotblamF$ for $b$ with respect to trading in competition with $a$. Using the Euler-Lagrange equation for $L_a$ and $L_b$ respectively yields the {\em equilibrium equations}:
    
    \begin{align}
        \ddot a &= -\frac{\lambda}{2} (\ddot b + \kappa \dot b)     \label{eq:eq-equation-ddota} \\
        \ddot b &= -\frac{1}{2\lambda} (\ddot a + \kappa \dot a)\label{eq:eq-equation-ddotb}
    \end{align}

    with the boundary conditions $a(0)=0, b(0)=0, a(1)=1, b(1)=1$. Since \cref{eq:eq-equation-ddota} and \cref{eq:eq-equation-ddotb} individually describe the conditions under which $a$ and $b$ are respectively paths that are minimal with respect to $L(a, \blam)$ and $L(\blam, a)$ respectively, they therefore represent being simultaneously the best response to one another. This is the essence of their being equilibrium strategies. The question is {\em how does one translate this to the quadratic programming setting?}

    The goal is to find a pair of strategies $(a, b_\lambda)$ that simultaneously minimizes $\CostF(a, \blam)$ and $\CostF(\blam, a)$ subject to a set of constraints on $a$ and $\blam$ respectively. We describe the algorithm now.

    \vskip 10pt

    {\bf Overview of the algorithm.} The algorithm may be described as replicating the adjustment dynamic process of traders repeatedly forming best-response strategies to their adversary's prior strategy. The entire process takes play in $\R^N$ where $N>$ is a predetermined number of Fourier coefficients used to approximate each cost function. Choose a number of Fourier coefficients $N$ and an initial guess $\blam^{(0)}\in \R^N$ representing $\blam$. Write $\ak\in \R^N$ and $\bk\in \R^N$ for respectively the values of Fourier coefficients in $a_\F$, $\blamF$ respectively after the $k$-th iteration of an iterative optimization. Write $\Cons(a; c)$ or $\Cons(b; d)$ for path or partial path constraints for strategy $a$ or $\blam$. Write $\CostF(a; \bk)$ to mean the cost function for $a$ with the value for the strategy $b$ fixed as the $k$-th guess, and similarly for $\Cost_\F(b; \ak)$. Then we have two separate minimization problems:

    \begin{center}
    \begin{tabular}{p{7.5cm}p{7.5cm}}
    {\begin{equation}
           \begin{array}{cl}
         \underset{(a_1\cdots a_N)}{\minimize} & \quad \CostF(a, \blam^{(k)}) \\[2.0ex]
         \text{subject to} & \quad \Cons(a; c)
         \label{opt:costab}
      \end{array}
      \end{equation}}
        & 
        {\begin{equation}
        \begin{array}{cl}
             \underset{(b_1\cdots b_N)}{\minimize} & \quad \CostF(\blam, a^{(k)}) \\[2.0ex]
            \text{subject to} & \quad \Cons(b; d)
            \label{opt:costba}
        \end{array}
        \end{equation}}
    \end{tabular}
    \end{center}
    
    The algorithm works by solving a pair of problems in each iteration. To start, in Step 1(i) we solve \cref{opt:costab} for $\Cost_\F(a, \blam^{(0)})$ which yields an $\argmin$ of $a^{(1)}$. Then in Step 1(ii) we solve \eqref{opt:costba} for $\Cost_\F(\blam, a^{(1)})$ to yield an $\argmin$ of $b^{(1)}$:

    \begin{equation}
        \begin{array}{lll}
           \textbf{Step 1(i):}  & \quad \text{Minimize $\Cost_\F(a, \blam^{(0)})$ s.t. $\Cons(a; c)$} &\quad  \textbf{{Produces}: $a^{(1)}$} \\
           \textbf{Step 1(ii):}  & \quad \text{Minimize $\Cost_\F(\blam, a^{(1)}$) s.t. $\Cons(b; d)$ } & \quad \textbf{Produces: $\blam^{(1)} $ }
        \end{array}
    \end{equation}

    Then the for $k>1$, the $k$-th step repeats:

    \begin{equation}
        \begin{array}{lll}
          \textbf{Step k(i):}  & \quad \text{Minimize $\CostF(a, \blam^{(k \shortminus 1)})$ s.t. $\Cons(a; c)$} &\quad  \textbf{{Produces}: $a^{(k)}$} \\
           \textbf{Step k(ii):}  & \quad \text{Minimize $\CostF(\blam, a^{(k)}$) s.t. $\Cons(b; d)$ } & \quad \textbf{Produces: $\blam^{(k)} $ }
        \end{array}
    \end{equation}

    The result of the optimization is a pair of points $(a^*, b^*)\in \R^N \times \R^N$ that correspond to the {\em reconstructed functions} which we identify by the same symbols:

    \begin{equation}
        \label{eq:reconstructed-fns}
        \begin{split}
            a^* &= (a_1, \dots, a_N) \mapsto a^*(t) = t + \asin \\
            b^* &= (b_1, \dots, b_N) \mapsto b^*(t) = t + \bsin
        \end{split}
    \end{equation}

    In practice we introduce a parameter $0<\gamma\le 1$ that allows one to step "part of the way" toward the new function at each step. We show the complete algorithm in \ref{alg:two-trader-eq} below.

    
    \begin{algorithm}[th]
    \caption{\bf (Alternating two-trader equilibrium)}
    \begin{algorithmic}[1]
    \STATE {\bf initialize:} $b^{(0)}, \Cons(a; c), \Cons(\blam; d)$, $k=1$, $N$, $0<\gamma< 1$.
    \WHILE{not converged} 
    \STATE {\bf Step k(i):}  \\
    \STATE \quad Compute: $a = \argmin_{(a_1\cdots a_N)} \CostF \big(a, \blam^{(k\shortminus 1)}\big) \st\, \Cons(a; c)$, see \eqref{opt:costab} \\
    \STATE $\quad \ak \leftarrow \gamma a + (1 \shortminus \gamma) a^{(k\shortminus 1)}$ \\
    \STATE  {\bf Step k(ii):} \\
    \STATE \quad Compute: $b = \argmin_{(b_1\cdots b_N)} \CostF \big(\blam, a^{(k)} \big)\, \st \, \Cons(\blam; d)$, see \eqref{opt:costba};
    \STATE \quad $\blam^{(k)} \leftarrow \gamma b + (1-\gamma) \blam^{(k\shortminus 1)} $
    \STATE Update: $k \leftarrow k+1$
    \ENDWHILE
    \STATE {\bf output:} $a^*=(a_1, \dots, a_N), b^*=(b_1, \dots, b_N)$; $a^*(t)=t + \asin$ and $b^*= t + \bsin$.
    \end{algorithmic}
    \label{alg:two-trader-eq}
    \end{algorithm}

    \vskip 10pt

    Note that in each successive step in \ref{alg:two-trader-eq} optimize by minimizing the cost function trading {\em not} versus the prior best-response strategy but rather versus a {\em convex combination} of the prior best-response and current best-response's Fourier coefficients. As we saw in Section \ref{sec:useful-facts-Fourier}, the convex combination of the strategy's Fourier coefficients is an approximation of the same convex combination of strategies.   We find that in practice this leads to significantly better convergence. In what follows we overload the term {\em best-response} to include these interpolations. See Figure \ref{fig:strats-state-space-change_gamma-2x5} for an illustration of the impact of $\gamma$ on the equilibrium path.

    \vskip 10pt

    {\bf Intuition.} Algorithm \ref{alg:two-trader-eq} has an intuitive interruption which explains why it will always converge to the two-trader equilibrium. After initialization, each step is simply the best-response to the other trader's best-response strategy. For example, to compute $a^{(k)}$ is to find the best-response to $\blam^{(k\shortminus 1)}$. The algorithm proceeds by successive best-responses to the last trader's best-response strategy until there is no more advantage to be gained, precisely the definition of equilibrium. Because this is a non-cooperative game, each trader can only react to the current best-response strategy as there is no available information on what either trader intends to do next. In Section \ref{sec:unconstrained-tte} we demonstrate that this can lead to cases where the cost of equilibrium are actually higher than non-equilibrium. We do this through the use of {\em state-space diagrams} and we use this to provide additional  insight into the nature of the equilibria we observe.

    \section{Numerical examples of two-trader equilibrium}

    In this section we provide numerical examples to illustrate using Algorithm \ref{alg:two-trader-eq} to calculate two-trader equilibria. As usual there is a unit trader with strategy $a$ and a $\lambda$-scaled trader with strategy $\blam$. We look at scenarios where $\kappa$ ranges from $1$ to $25$ with various values for $\lambda$. In the first part below we look at unconstrained two-trader equilibria. In this part we compare the output of Algorithm \ref{alg:two-trader-eq} to the results given by the closed-form solutions in \eqref{eq:two-trader-eq-a} and \eqref{eq:two-trader-eq-b}. 

    \subsection{Computing equilibria and the dynamic path to equilibrium} 
    \label{sec:unconstrained-tte}
    
    We begin by studying two-trader equilibria with {\em no constraints} to achieve two purposes. With no constraints we can compare the optimization results to the closed form solutions in \eqref{eq:two-trader-eq-a} and \eqref{eq:two-trader-eq-b}; and we can study the {\em path} to convergence build up from each step in Algorithm \ref{alg:two-trader-eq}. This provides insight into the nature of the dynamic process of adjustments that leads to convergence.

    To provide a measure of the quality of the results, we use the $L_2$ norm of the difference between the Fourier approximation and the closed form solution. For $\kappa$ and $\lambda$ let $(a_1,\dots, a_N)$ be the Fourier coefficients of the result of Algorithm \ref{alg:two-trader-eq} for strategy $a$. Then let $a^*(t) = t + \asin$ be the corresponding {\em reconstructed} function; and let $\aeq(t)$ be the two-trader equilibrium strategy for the same $\kappa, \lambda$ in \cref{eq:two-trader-eq-a}. Write $\|a^* - \aeq\|_{2}$ for the $L_2$ norm of the difference between $a^*$ and $\aeq$:

    \begin{equation}
        \|a^* - \aeq\|_{2} = \left( \intzo \left(a^*(t) - \aeq(t)\right)^2 \dt \right)^{1/2}
    \end{equation}

    We will use this as our measure of the quality of the solution $a^*$ versus the closed-form solution $\aeq$. Similarly we can form $b^*$ and use $\|b^* - \beq\|_{2}$ to measure the quality of the solution of $b^*$. 

    \vskip 10pt
    
\begin{figure}[h!]
        \centering
        \includegraphics[width=0.75\textwidth, height=0.5\textwidth]{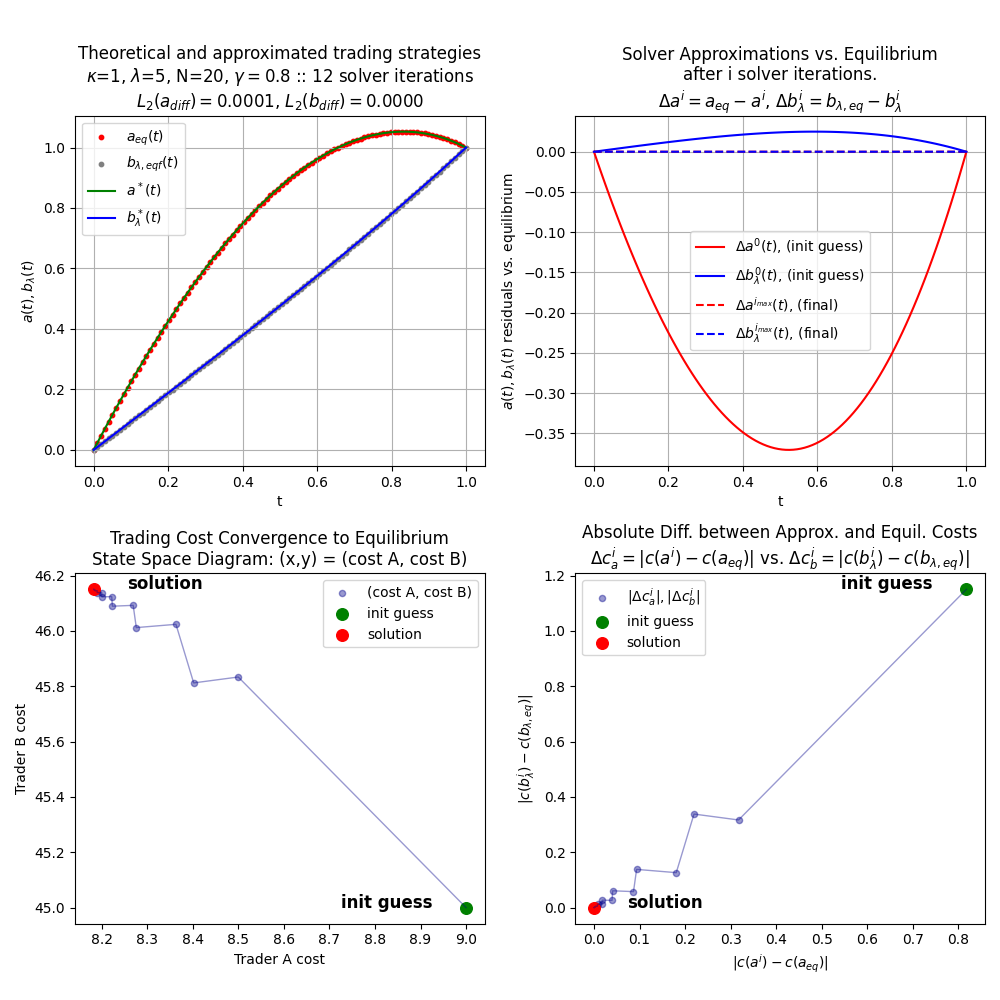}
        \caption{Two-trader equilibria strategies $a^*, \blam^* $ compared with the closed-form solutions $\aeq, \beq$ in \eqref{eq:two-trader-eq-a} and \eqref{eq:two-trader-eq-b} for $\kappa=1$ and $\lambda=5$ using $N=20$ Fourier terms. The upper left-hand plot shows $a^*$ traded along with $\aeq$ and $b^*$ along with $\beq$, as well as the $L_2$ norms of the differences between the theoretical equilibrium strategies and their approximations. The upper right-hand shows the difference between the reconstructed functions and theoretical equilibrium functions at the first and last iterations of the optimization. The state-space diagram in the lower-left plot shows the total cost of trading for $a$ and $\blam$ directly prior to Step 1(i) when both traders trade a risk-neutral strategy. The plot shows that the solution (the equilibrium strategy)  benefits $a$, lowering the total cost trading from approximately 9 to 8.2, while increasing $\blam$'s modestly 45 to approximately 46.2. The solver converged in 12 iterations and $\gamma=0.8$.}
        \label{fig:tte-k1-l5}
    \end{figure}

\begin{figure}[h!]
        \centering
        \includegraphics[width=0.85\textwidth, height=0.45\textwidth]{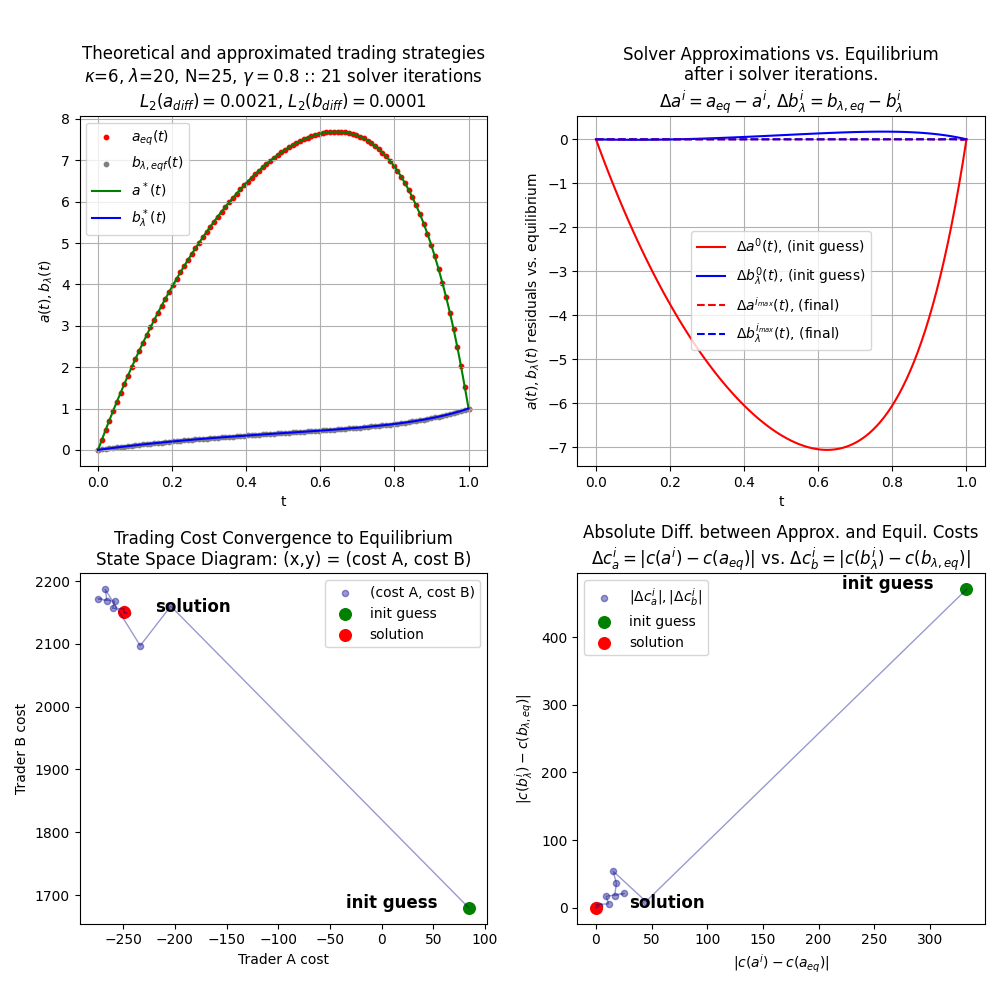}
        \caption{Two-trader equilibria solutions compared with the closed-form for $\lambda=20, \kappa=6$ and $N=25$ Fourier series terms for Algorithm \ref{alg:two-trader-eq} with $\gamma=0.8$ and Algorithm \ref{alg:two-trader-eq} converged in 21 iterations. The solutions presented are the reconstructed functions $a^*$ and $b^*$ in \cref{eq:reconstructed-fns} and are compared with the closed-form solutions $\aeq$ and $\beq$ in equations \eqref{eq:two-trader-eq-a} and \eqref{eq:two-trader-eq-b}. The top left chart shows the $L_2$ norms of the differences between the theoretical equilibrium strategies and their approximations.. The state-space diagram plots the cost of trading for $a$ and $\blam$ starting with the initial point (labeled {\em init}) which represents both traders trading the risk-neutral strategy. In this case the $\blam$ trades 20 times as much as $a$ and $a$'s initial best-response to $\blam=\lambda\cdot t$ already substantially reduces its cost while increasing $\blam$'s.}
        \label{fig:tte_k6_l20}
    \end{figure}

\begin{figure}[h!]
        \centering
        \includegraphics[width=0.85\textwidth, height=0.45\textwidth]{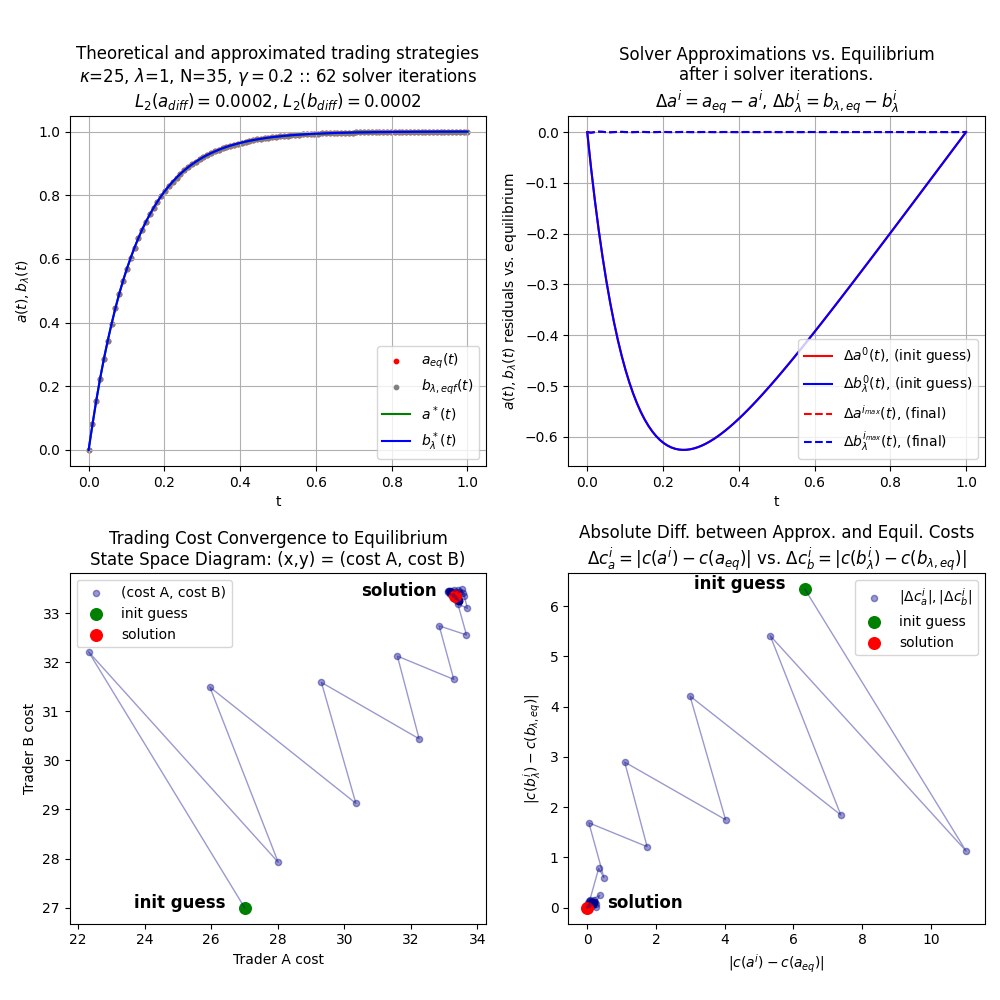}
        \caption{Unconstrained two-trader equilibria solutions compared with the closed-form for $\lambda=1, \kappa=25$ and $N=35$ terms in the Fourier series approximations for Algorithm \ref{alg:two-trader-eq} with $\gamma=0.2$ and the algorithm converged in 62 iterations. The solutions presented are the reconstructed functions $a^*$ and $b^*$ in \cref{eq:reconstructed-fns} and are compared with the closed-form solutions $\aeq$ and $\beq$ in equations \eqref{eq:two-trader-eq-a} and \eqref{eq:two-trader-eq-b}. The top left chart shows the $L_2$ norms of the differences between the theoretical equilibrium strategies and their approximations. The state-space diagram plots the cost of trading for $a$ and $\blam$ starting with the initial point (labeled {\em init}) which represents both traders trading the risk-neutral strategy. Of interest is that the total cost of trading is higher for both traders at equilibrium than it would be if they both traded a risk-neutral strategy as they do at the point labeled {\em init guess}. That is, they would both be better off if they could coordinate.  It's also interesting that the convergence is much slower for higher $\gamma$, for example if we use the same value of $\gamma=0.8$ as in the previous example, the solver does converge within tolerance even after 100 iterations.}
        \label{fig:tte_k25_l1}
    \end{figure}

    {\bf Plots:} Figures \ref{fig:tte-k1-l5} to \ref{fig:tte_k25_l1} provide three illustrations of solving the unconstrained two-trader equilibrium using Algorithm \ref{alg:two-trader-eq}. Each optimization starts with the initial guess $a^{(0)}=t$, the risk-neutral strategy and the first iteration is $\blam^{(1)}$ the best-response to this strategy. Unless otherwise stated, these plots $\gamma=0.8$ (see \ref{alg:two-trader-eq}). We use a lower value when convergence is an issue. The upper left-hand plot of each shows the trading strategies for $a$ and $\blam$ on top of one another; the upper right-hand plot shows the difference between the closed-form and approximate solutions for both the initial iteration and the final iteration of the optimization. For example, in Figure \ref{fig:tte-k1-l5} ($\lambda=5$ and $\kappa=1$) the right hand plot shows in solid red the plot of $b^{(1)}(t) - \beq(t)$ which shows that at approximately $t=0.5$ the equilibrium strategy strategy holds approximate 0.35 units more than the initial guess (the risk-neutral strategy). The initial guess for $a$ is a lot closer in absolute terms (dark blue) but with $\lambda=5$ this scale difference makes sense.

    \vskip 10pt

    {\bf Equilibrium paths and consequences of no-collusion:} As noted earlier Algorithm \ref{alg:two-trader-eq} represents the dynamic path to the equilibrium strategy, showing successive best-responses to the prior best-response strategy. In cases where two traders are in competition and trading {\em similar} quantities of stock the results reveal an important consequence of prohibitions on {\em collusion}. The conclusion is that in certain instances where two traders are trading approximately the same quantity, the costs to each is lower when they both trade the risk-neutral than when they trade the equilibrium strategy. However, because the traders cannot coordinate their trading they each must successively fashion the best-response to the other's best response and inevitably pay more. This is impossible to avoid without some form of collusion.
    
\begin{figure}[h!]
        \centering
        \includegraphics[width=0.85\linewidth]
        {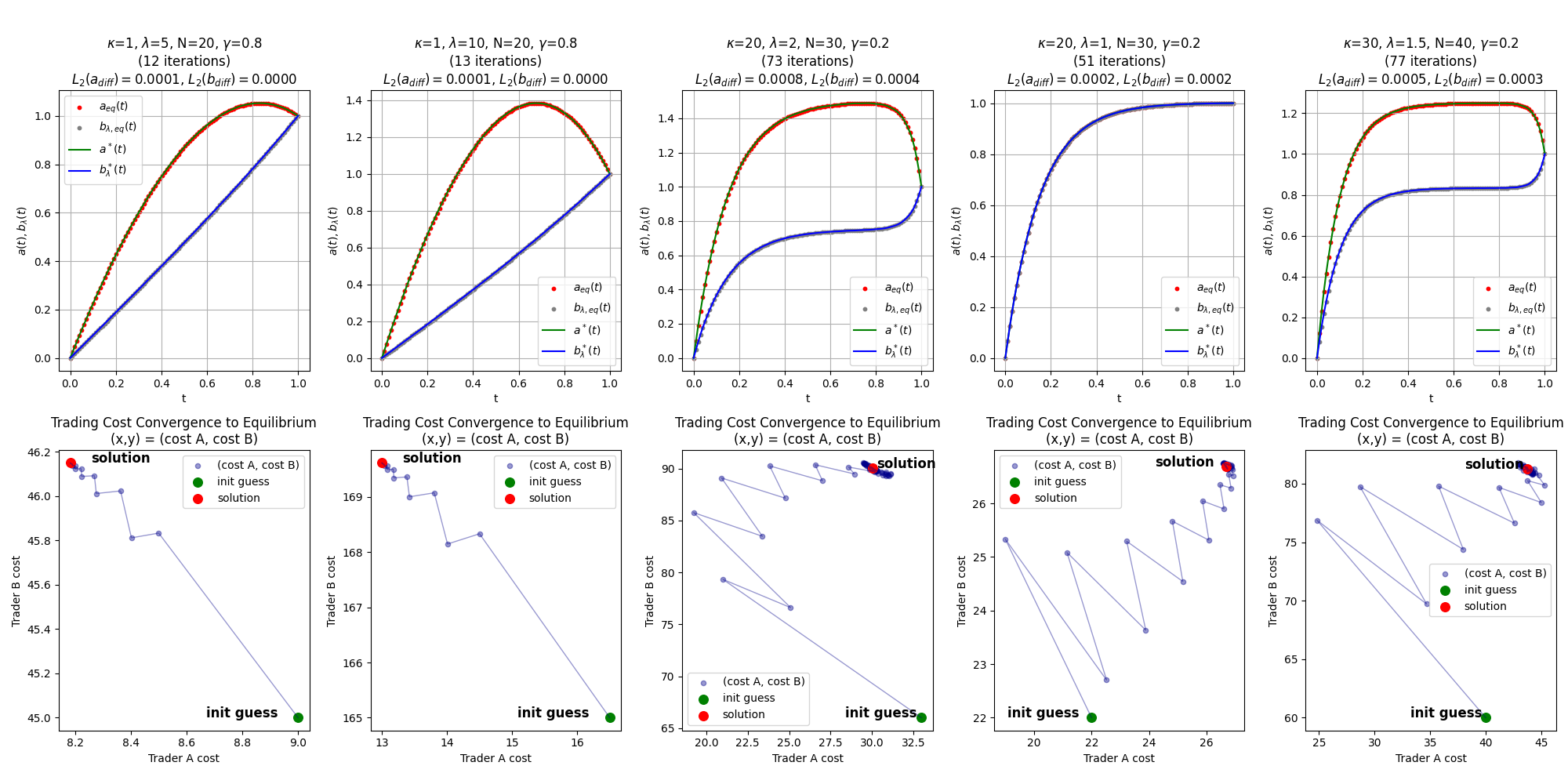}
        \caption{Unconstrained two-trader equilibria solutions compared with the closed-form for a cross-section of parameters using Algorithm \ref{alg:two-trader-eq} with the values of $\gamma$ selected to ensure quick convergence (i.e. $\gamma=0.8$ for the left two plots, and $\gamma=0.2$ for the right three). The solutions presented are the reconstructed functions $a^*$ and $b^*$ in \cref{eq:reconstructed-fns} and are compared with the closed-form solutions $\aeq$ and $\beq$ in equations \eqref{eq:two-trader-eq-a} and \eqref{eq:two-trader-eq-b}. The number of terms $N$ for each set of $\lambda$ and $\kappa$ is chosen to ensure a given level of approximation accuracy, which we measure by the $L_2$ norm of the difference between the exact analytic and the approximated trading strategy (shown above each top chart).  The state-space diagram in the bottom row plots the cost of trading for $a$ and $\blam$ starting with the initial point (labelled {\em init guess}), which represents both traders trading the risk-neutral strategy. Of interest is for the two rightmost charts the total cost of trading is higher for both traders than it would be if they both traded a risk-neutral strategy as in initiation -- they would both be better off, but because collusion is prohibited, they inevitably both end up trading the equilibrium strategies, which are more expensive than trading the risk-neutral strategies.}
        \label{fig:strats-state-space-2x5}
    \end{figure}

    We can see this in Figure \ref{fig:tte_k25_l1} which displays data for the unconstrained two-trader equilibrium for the case where $\lambda=1$ and $\kappa=5$ and specifically $\gamma=0.2$. The lower left-hand plot shows the state-space diagram for the optimization. The axes of the plot are the cost of trading for $a$ and $b$ respectively and it plots $(a^{(k)}, b^{(k\shortminus 1)})$, the result of step k(i), and $(a^{(k)}, b^{(k)})$, the result of step k(ii). The point labeled {\em init guess} shows the costs of the pre-initialization strategies, that is, when both $a$ and $b$ trade the risk-neutral strategy. Both costs are approximately 27. The initial guess is connected by a line to a second point whose coordinates are approximately $(22, 32)$. This point represents the pair of strategies consisting of $a$, which is a convex combination of the best-response to $b$ trading the risk-neutral strategy and the risk-neutral strategy (see the use of $\gamma$ in Algorithm \ref{alg:two-trader-eq}). Clearly this step {\em decreases} $a$'s total cost of trading while it {\em increases} $b$'s. However, the next iteration, which is $b$ trading the best response to $a$ from the prior iteration, decreases the total cost of trading for $b$ while increasing the cost for $a$. And it exactly as such. The net result is that {\em both traders} end up paying more than simply trading the risk-neutral strategies. An obvious conclusion is that if the two-traders were allowed to collude they would be able to conclude that both trading the risk-neutral strategy would be mutually beneficial but as with many two-person non-cooperative games, such as prisoners dilemma. A similar phenomena is also show in the right two plots in Figure \ref{fig:strats-state-space-2x5}.

    \clearpage
    \subsection{Plots of unconstrained two-trader equilibrium}

    In the following plots we examine two-trader equilibria for various values of $\kappa$ and $\lambda$ with $\gamma=0.2$ in all cases. At this stage we have no investigated what the optimal value for $\gamma$ is but illustrate in the sequel that the algorithm proposed has instances that converge with $\gamma<1$ but not with $\gamma=1$.

    \vskip 10pt

\begin{figure}[h!]
        \centering
        \includegraphics[width=0.85\linewidth]
        {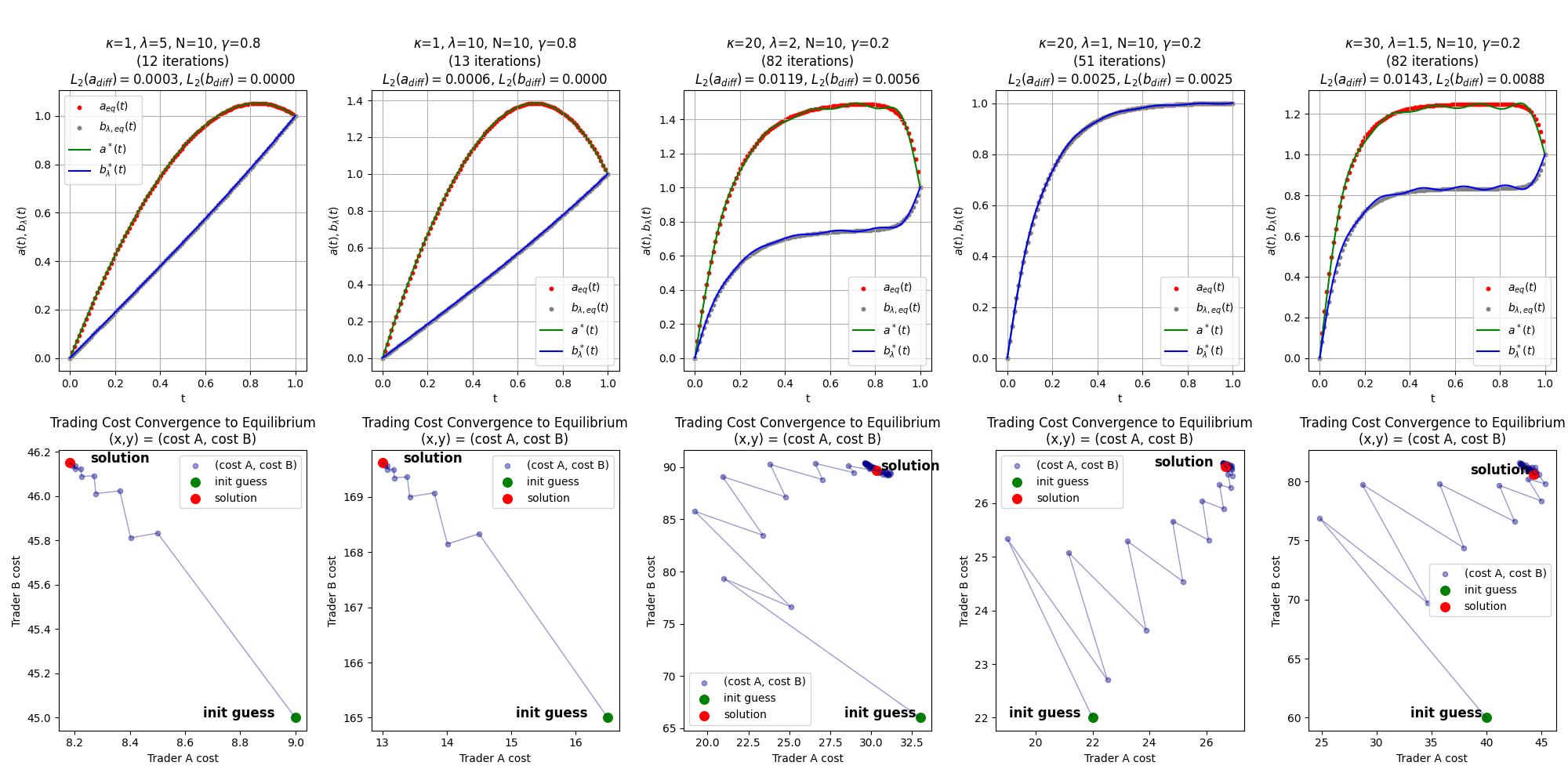}
        \caption{Unconstrained two-trader equilibria solutions compared with the closed-form for a cross-section of parameters using Algorithm \ref{alg:two-trader-eq} with the values of $\gamma$ selected to ensure quick convergence (i.e. $\gamma=0.8$ for the left two plots, and $\gamma=0.2$ for the right three). The sets of parameters $\lambda$, $\kappa$, $\gamma$ is the same as in the previous chart \ref{fig:strats-state-space-2x5}, but the number of Fourier terms is fixed at lower $N=10$ for all charts.  Both the $L_2$ norms of the approximation error and visual inspection illustrate the dependence of the approximation accuracy on the number of Fourier terms $N$.  For example, for $\kappa=20$, $\lambda=1$ parameters (second chart from the right) the $L_2$ approximation error for $N=10$ is almost 12 times higher than for $N=30$ (0.0025 vs. 0.0002).}
        \label{fig:strats-state-space-n10-2x5}
    \end{figure}

\begin{figure}[h!]
        \centering
        \includegraphics[width=0.85\linewidth]{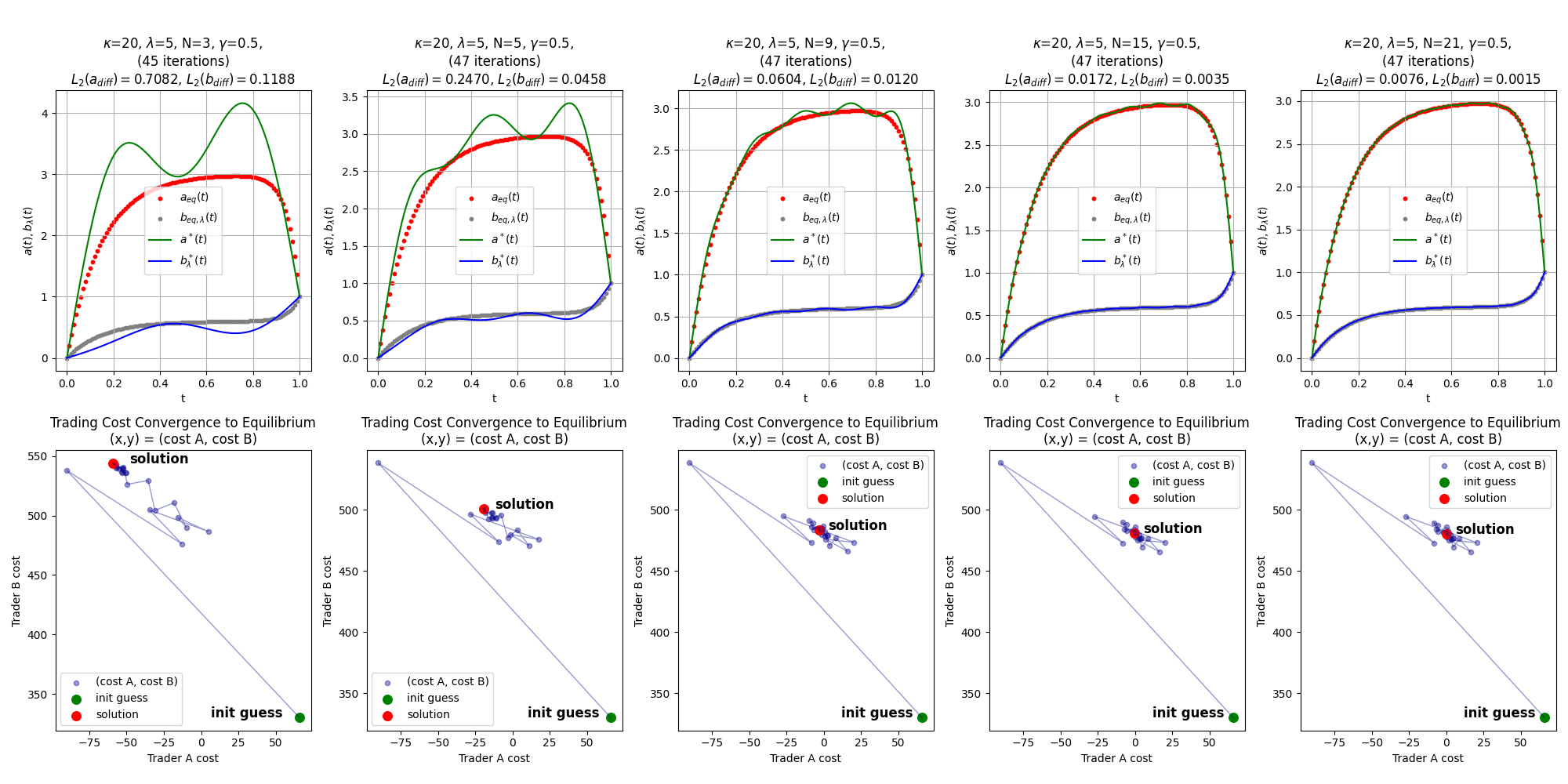}
        \caption{This plot demonstrates the impact of a the number of Fourier coefficients $N$ used in forming the approximate cost functions (see Propositions \ref{prop:approx-cost-ab} and \ref{prop:approx-cost-ba}) for the case $\kappa=20, \lambda=5$ and $\gamma=0.5$.}
        \label{fig:strats-state-space-change_N-2x5}
    \end{figure}

\begin{figure}[h!]
        \centering
        \includegraphics[width=0.85\linewidth]{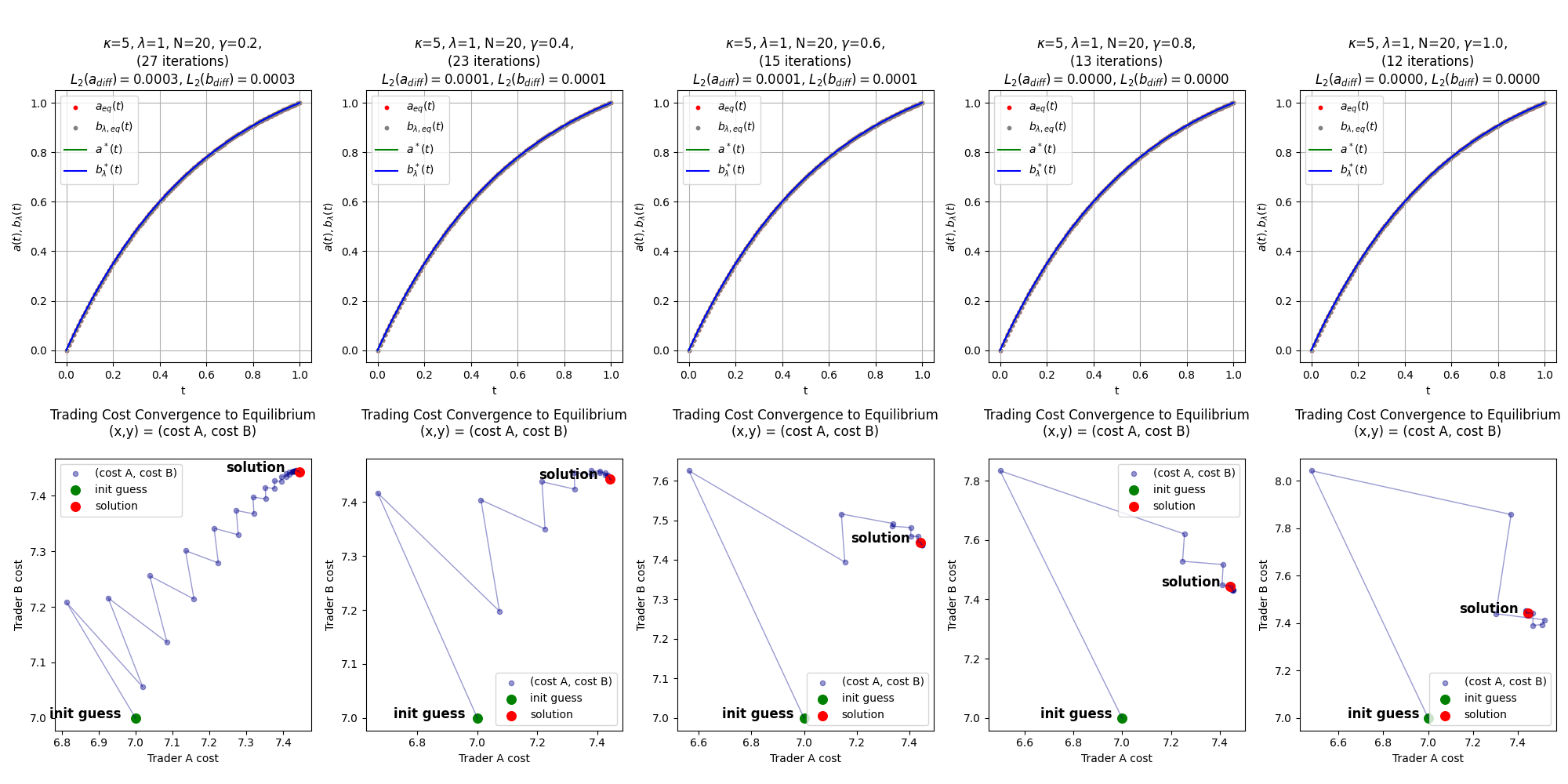}
        \caption{Unconstrained two-trader equilibrium for $\kappa=5$, $\lambda=1$ for $\gamma$ increasing from left to right from 0.2 to 1.0; higher $\gamma$ decreases the number of iterations required for convergence, while higher $\gamma$ eventually causes the solver to diverge.}
        \label{fig:strats-state-space-change_gamma-2x5}
    \end{figure}

\begin{figure}[h!]
        \centering
        \includegraphics[width=0.85\linewidth]{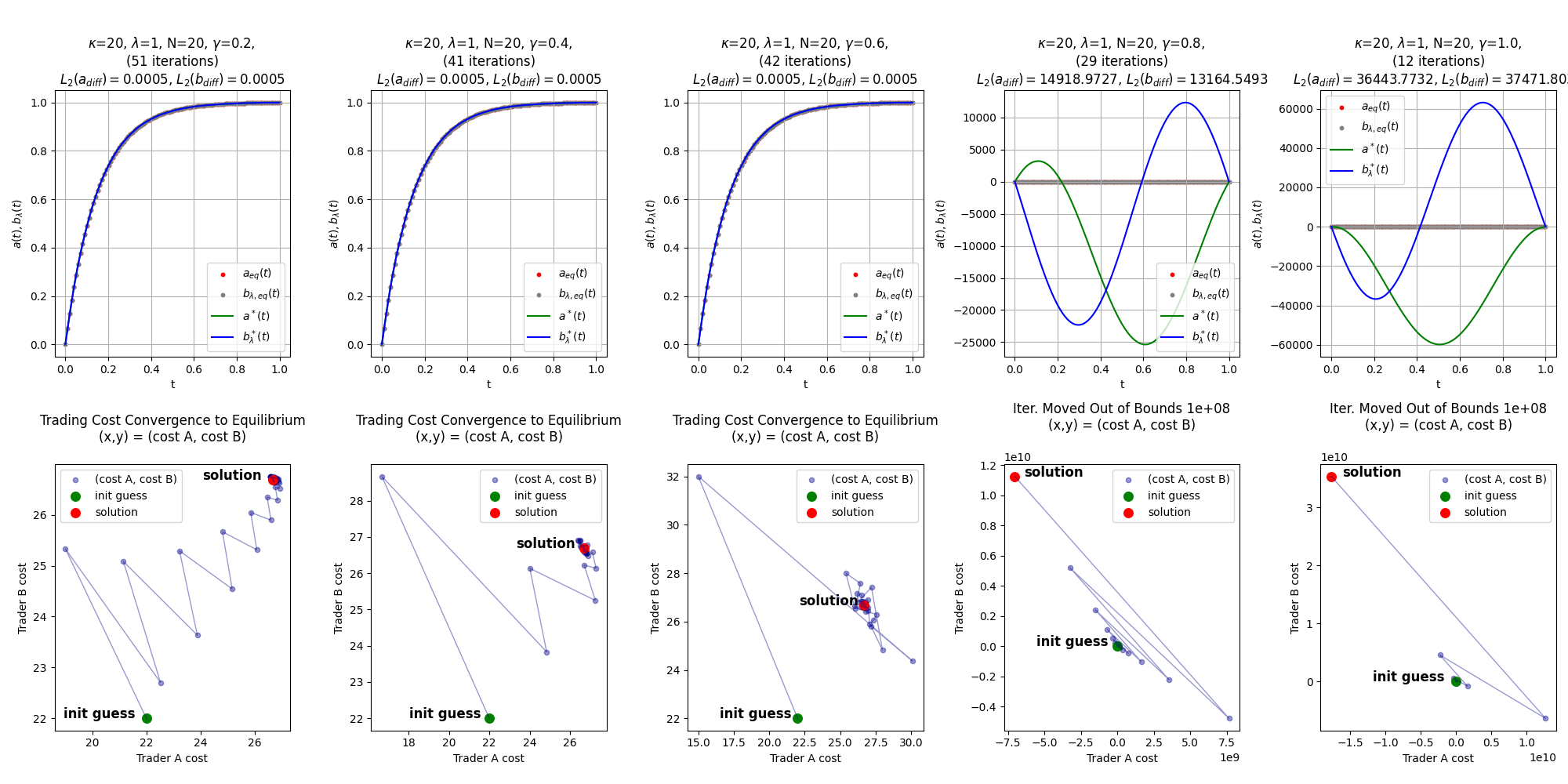}
        \caption{A case where Algorithm \ref{alg:two-trader-eq} diverges when $\gamma$ is too large.}
        \label{fig:strats-state-space-change_gamma-diverge-2x5}
    \end{figure}

\begin{figure}[h!]
        \centering
        \includegraphics[width=0.8\linewidth]
        {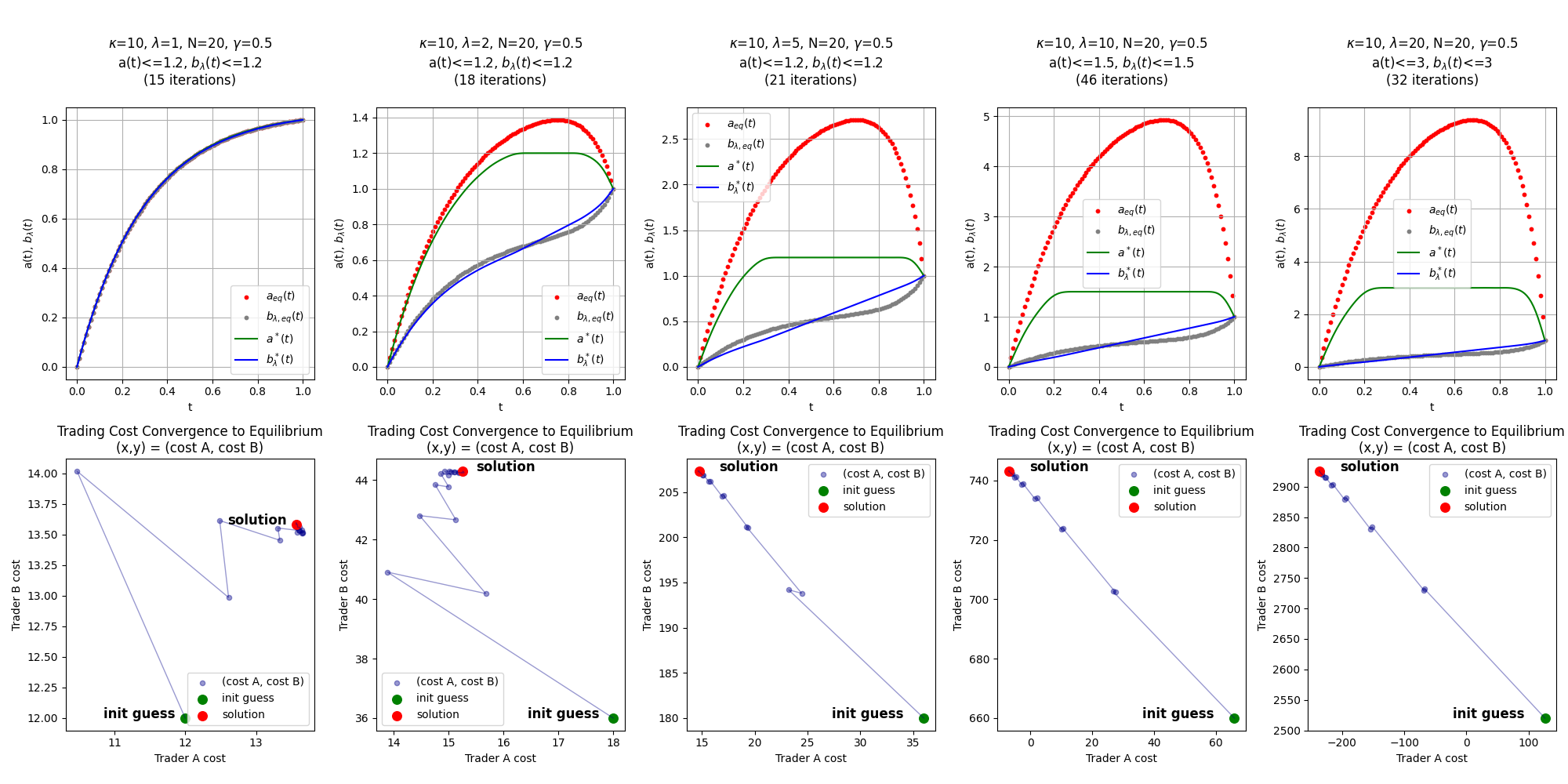}
        \caption{Two-trader constrained equilibria solutions compared with the unconstrained equilibrium closed-form for a cross-section of parameters using Algorithm \ref{alg:two-trader-eq} with $\gamma=0.2$. The solutions presented are the reconstructed functions $a^*$ and $b^*_{\lambda}$ in \cref{eq:reconstructed-fns}, representing constrained equilibrium trading strategies, subject to constraints on overbuying and short selling, see \ref{sec:constraint-overbuy} and \ref{sec:constraint-short-selling}.  For reference the chart also includes  the closed-form solutions $\aeq$ and $\beq$ in equations \eqref{eq:two-trader-eq-a} and \eqref{eq:two-trader-eq-b}. The number of terms $N$ for each set of $\lambda$ and $\kappa$ is chosen to ensure high approximation accuracy.  The state-space diagram in the bottom row plots the cost of trading for $a$ and $\blam$ starting with the initial point (labelled {\em init guess}), which represents both traders trading the risk-neutral strategy.  As the relative size of trader B, $\lambda$, increases, trader A tries to build up progressively greater positions to benefit from the permanent market impact, $\kappa$, until the overbuy constraint is hit.  Of interest is that for the left column of charts the total cost of trading is higher for both traders than it would be if they both traded a risk-neutral strategy as in initiation -- they would both be better off, but because collusion is prohibited, they inevitably both end up trading the constrained equilibrium strategies, which are more expensive than trading the risk-neutral strategies.  In all other columns, the smaller trader A reduces her trading costs at the expense of trader B, whose costs increase.}
        \label{fig:strats-state-space-cons-kapp10-2x5}
    \end{figure}

    {\bf Ensemble plots:} In Figures \ref{fig:strats-state-space-n10-2x5} to \ref{fig:strats-state-space-cons-kapp10-2x5} we depict "ensembles" of plots of two-trader equilibrium to demonstrate how a single parameter's change influences the optimization results. For example, as explained in Algorithm \ref{alg:two-trader-eq} it can be helpful to advance in each step using convex combinations of prior- and next-step strategies. We have not explored how to optimally set $\gamma$ but in Figure \ref{fig:strats-state-space-change_gamma-diverge-2x5} we demonstrate for the case $\kappa=25$ and $\lambda=1$ the algorithm diverges when $\gamma=1$ and converges when $\gamma\le 0.6$.

    \vskip 10pt

    \subsection{Two-trader equilibria with constraints}

    As a final set of numerical demonstrations we calculate two-trader equilibria with constraints using Algorithm \ref{alg:two-trader-eq} to calculate two-trader equilibria with various constraints.

\clearpage
\appendix
\section{Proofs of cost function approximations}

    For completeness this section proves the expressions for the cost function approximations given in Propositions \ref{prop:approx-cost-ab} and \ref{prop:approx-cost-ba}. These proofs are largely a consequence of the trigonometric identities in Section \ref{sec:key-trig-identities} and basic calculus. 

    \subsection{Proof of Proposition \ref{prop:approx-cost-ab}}
    \label{sec:proof-cost-fn-approx-ab}

    To prove this we proceed by noting the total cost function may be expressed as the sum of four integrals by expanding the terms of the loss function $(\dot a + \lambda\dotblam)\dot a + \kappa (a + \lambda \blam) \dot a$:

    \begin{align}
        \Costab &= \int_0^1 (\dot a + \lambda\dot \blam)\dot a + \kappa (a + \lambda \blam) \dot a \dt \\
                &= \int_0^1 \dot a \dot a \dt + \int_0^1 \lambda \,\dot \blam \dot a \dt +
                        \int_0^1 \kappa \, a \dot a \dt + \int_0^1 \kappa \lambda \, \blam \dot a \dt\\
                &= \int_I + \int_{II} + \int_{III} + \int_{IV}          
    \end{align}

    where each term is determined by the appropriate Fourier expansions:
    
    {\bf The first term $\int_I = \int_0^1 \dot a \dot a \dt$:} We substitute $\dot a_\F$ for $\dot a$
    
    \begin{align}
    \label{eq:int_I}
    \int_I &= \int_0^1 \dot a_\F \dot a_F \dt\\
           &= \int_0^1 \left(1+\acos\right)\left(1+\acos\right) \dt \\
           &= \int_0^1 1 + 2\acos + \Big( \acos \Big)^{\! 2} \dt \\ 
           &= \int_0^1 1 \dt + 2 \sum_{n=1}^N a_n \int_0^1 n \pi \cos(n \pi t) \dt 
                +  \sum_{n=1}^N \sum_{m=1}^N \int_0^1 a_n a_m n^2 m^2 \pi^2 \cos(n\pi t) \cos(m\pi t) \dt \\
           &= 1 + \frac{\pi^2}{2} \sum_{n=1}^{N} n^2  a_n^2
    \end{align}

    The final equality follows from the identities \eqref{eq:fourier-b-func} and the trigonometric integral identity \eqref{eq:key-trig-c}, respectively. Summarizing:

    \begin{equation}
        \label{eq:int_I_final}
        \int_0^1 \dot a_\F \dot a_F \dt = 1 + \frac{\pi^2}{2} \sum_{n=1}^{N} n^2  a_n^2 
    \end{equation}

    \vskip 10pt

    {\bf The second term $\int_{II}$ is} given by $\int_0^1 \lambda \,\dot b \dot a \dt$ and substituting $\dot a_\F, \dotblamF$ for $a$ and $b$ respectively:
    
    \begin{align}
    \label{eq:int_II}
        \int_{II}   &= \int_0^1 \lambda\, \dotblamF \dot a_\F \dt \\
                    &= \lambda \int_0^1 \Big( 1 + \bcos \Big) \Big(1 + \acos \Big) \dt \\
                    &= \lambda \int_0^1 1 + \bcos + \acos + \bcos\acos  \dt \\
                    &= \int_0^1 1 \dt + 2 \sum_{n=1}^N (b_n + a_n)\int_0^1 n \pi \cos(n \pi t) \dt 
                        + \sum_{n=1}^N \sum_{m=1}^N \int_0^1 b_n a_m n^2 m^2 \pi^2 \cos(n\pi t) \cos(m\pi t) \dt \\
                    &= \lambda + \lambda \frac{\pi^2}{2} \sum_{n=1}^{N} n^2 a_n b_n
    \end{align}

    Where the final equality follows from the identities \eqref{eq:key-trig-b} and \eqref{eq:key-trig-d} of Section \ref{sec:key-trig-identities}, respectively. Summarizing:

    \begin{equation}
        \label{eq:int_II_final}
        \int_0^1 \lambda\, \dotblamF \dot a_\F \dt = \lambda + \lambda \frac{\pi^2}{2} \sum_{n=1}^{N} n^2 a_n b_n 
    \end{equation}

    \vskip 10pt

    {\bf The third term $\int_{III}$} corresponds to the integral $\int_0^1 \kappa a(t) \dot{a}(t) dt$ and again is obtained substituting $\dot a_\F$ for $\dot a$ and $a_\F$ for $a$.

    \begin{subequations}
    \begin{align}
        \int_{III}  &= \int_0^1 \kappa \,\dot a_\F a_\F \dt \label{eq:int_III_a}\\
                    &= \int_0^1 \kappa\, \Big(1 + \acos\Big)\Big(t + \asin\Big) \dt\label{eq:int_III_b} \\
                    &= \kappa \int_0^1 t + \asin + t \cdot \acos + \acos \asin \dt     \label{eq:int_III_c}\\
                    &= \kappa \!\! \left( \int_0^1 t \dt
                        +  \!\!\intasin +  \!\!\intacos + \!\!\int_0^1 \sum_{m, n=1}^N a_n a_m n \pi \cos(n \pi t) \sin(m \pi t) \dt\right)   \label{eq:int_III_d}
    \end{align}        
    \end{subequations}

    To integrate the last term \cref{eq:int_III_d} we use the identity \cref{eq:int_cos_sin} from Section \ref{sec:key-trig-identities}:

    \begin{align}
        \kappa\int_0^1 \sum_{m, n=1}^N a_n a_m n \pi \cos(n \pi t) \sin(m \pi t) \dt 
                &= \kappa \! \! \sum_{n, m=1}^N \int_0^1 a_n a_m n \pi \cos(n \pi t) \sin(m \pi t) \dt \\
                &= \kappa \!\!\! \sumnmodd a_n a_m n \pi \frac{2m}{\pi(m^2 - n^2)}\\
                &= 2\kappa \bigbs \sumnmodd  a_n a_m n \frac{m}{m^2 - n^2}
    \end{align}

    and we note that this term is zero because for every pair $(m,n), n+m \text{odd}$, the is a corresponding pair $(m,n)$ which cancels. This leads to \cref{eq:int_III_d} becoming
    
    \begin{align}
        \frac{\kappa}{2} -\kappa \sumodd \frac{2 a_n}{n \pi} + \kappa \sumodd \frac{2 a_n}{n \pi} + 
                    2 \kappa\bigbs \sumnmodd n m \cdot \frac{a_n a_m}{m^2 - n^2}
            &= \frac{\kappa}{2} 
    \end{align}

    Summarizing we have the following expression for $\int_{III}$:

    \begin{equation}
        \label{eq:int_III_final}
        \int_0^1 \kappa \,\dot a_\F a_\F = \frac{\kappa}{2} 
    \end{equation}

    {\bf The fourth term $\int_{IV}$} corresponds to $\int_0^1 \kappa \lambda \, b \dot a \dt$ and we proceed by substituting $\blamF$ for $b$ and $a_\F$ for $a$:

    \begin{align}
    \label{eq:int_IV}
    \int_{IV} &= \int_0^1 \kappa \lambda \blamF \dot a_\F \dt \\
              &= \kappa \lambda \int_0^1 \Big(t + \bsin \Big) \Big(1 + \acos\Big) \dt \\
              &= \kappa \lambda \int_0^1 t + t \acos + \bsin + \bsin \acos \dt 
    \end{align}

    To integrate the last term we proceed analogously to what we did with $\int_{III}$:

    \begin{align}
        \kappa \lambda \int_0^1 \bsin \acos\dt &= \kappa \lambda \sum_{n, m=1}^N \int_0^1 a_n b_m n \pi \cos(n\pi t) \sin(m \pi t) \dt \\
            &= 2\kappa \lambda \!\!\!\sumnmodd a_n b_m n \frac{m}{m^2 - n^2}
    \end{align}

    We can see that the second and third terms of this integral do not cancel and instead become sums of terms $\frac{2(b_n-a_n)}{n \pi}$ and so the final expression for $\int_{IV}$ is as follows:

    \begin{equation}
        \label{eq:int_IV_final}
        \int_0^1 \kappa \lambda \blamF \dot a_\F \dt =
                \kappa\lambda  \left( \frac{1}{2} + \sumodd \frac{2(b_n- a_n)}{n \pi} + 2\!\!\! \sumnmodd a_n b_m  n \cdot \frac{a_m b_n}{m^2 - n^2} \right)
    \end{equation}

    {\bf Computing the cost function as the sum $\int_{I} + \int_{II}+\int_{III} + \int_{IV}$:} Finally we arrive at the cost function via summing the previous results. Summarizing we have:

    \begin{align}
        \int_I &= 1 + \frac{\pi^2}{2} \sum_{n=1}^{N} n^2  a_n^2 \\
        \int_{II}   &= \lambda + \lambda \frac{\pi^2}{2} \sum_{n=1}^{N} n^2 a_n b_n\\
        \int_{III}  &= \frac{\kappa}{2} + 2 \kappa \bigbs\sumnmodd  a_n a_m n \frac{m }{m^2 - n^2}\\
        \int_{IV}   &=  \kappa\lambda \left( \frac{1}{2} + \sumodd \frac{2(b_n- a_n)}{n \pi} + 2 \bigbs\sumnmodd a_n b_m  n \cdot \frac{m}{m^2 - n^2} \right)
    \end{align}

    Collecting the constant terms we have:

    \begin{align}
        1 + \lambda + \frac{\kappa}{2} + \frac{\kappa\lambda}{2} &= 1 + \lambda + \frac{\kappa(1+\lambda \blam)}{2}\\
                       &= \frac{(2+\kappa)(1+\lambda \blam)}{2}
    \end{align}

    Next combining the non-constant terms of integrals I and II and the first sum of integral IV we obtain:

    \begin{equation}
        \frac{\pi^2}{2} \sum_{n=1}^N n^2 (a_n^2 + \lambda a_n b_n) + \kappa\lambda \sumodd \frac{2(b_n- a_n)}{n \pi}
    \end{equation}

    and finally combining the summation in integral III and the last summation integral IV we have:

    \begin{equation}
        2\kappa\lambda\!\!\! \sumnmodd a_n b_m  n \cdot \frac{m} {m^2 - n^2} 
    \end{equation}

    Putting the three terms together we arrive at:

    \begin{equation}
        \CostFab= \frac{(2+\kappa)(1+\lambda)}{2} + 
                    \frac{\pi^2}{2} \sum_{n=1}^N n^2 (a_n^2 + \lambda a_n b_n) + \kappa\lambda \sumodd \frac{2(b_n- a_n)}{n \pi} +
                    2\kappa \bigbs \sumnmodd \frac{\lambda b_m  a_n \, n \, m}{m^2 - n^2}
    \end{equation}

    Rearranging the terms we obtain the proof of Proposition \ref{prop:approx-cost-ab}. 

    \subsection{Proof of Proposition \ref{prop:approx-cost-ba}}
    \label{sec:proof-cost-fn-approx-ba}

    The derivation of the formula proceeds almost identically with the prior one as follows:

    \begin{align}
        \CostFba &= \intzo (\dot a_\F + \lambda \dotblamF)\lambda\dotblamF + \kappa(a_\F + \lambda \blamF)\lambda \dotblamF \dt \\
            &= \lambda \intzo \dot a_\F \dotblamF \dt + 
                \lambda^2 \intzo \dotblamF \dotblamF \dt + 
                \kappa \lambda \intzo a_\F\dotblamF \dt  + 
                \kappa \lambda^2 \intzo \blamF\dotblamF \dt \\ 
            &= \int_I^{ba} + \int_{II}^{ba} + \int_{III}^{ba} + \int_{IV}^{ba}
    \end{align}

    Recall:

    \begin{subequations}
    \begin{align}
        \int_0^1 \dot a_\F \dot a_F \dt &= 
                1 + \frac{\pi^2}{2} \sum_{n=1}^{N} n^2  a_n^2 \label{eq:int_I_ba}\\
        \lambda \int_0^1 \, \dotblamF \dot a_\F \dt &= 
                \lambda + \lambda \frac{\pi^2}{2} \sum_{n=1}^{N} n^2 a_n b_n \label{eq:int_II_ba}\\
        \kappa \int_0^1 \,\dot a_\F a_\F \dt &= 
                \frac{\kappa}{2} + 2 \kappa \bigbs\!\sumnmodd  a_n a_m n \frac{m }{m^2 - n^2} \label{eq:int_III_ba} \\
        \kappa \lambda \int_0^1  \blamF \dot a_\F \dt &=
                \kappa\lambda \left( \frac{1}{2} + \sumodd \frac{2(b_n- a_n)}{n \pi} + 2 \bigbs\sumnmodd a_n b_m  n \cdot \frac{m}{m^2 - n^2} \right) \label{eq:int_IV_ba}
    \end{align}
    \end{subequations}

    With the above identities we can provide formulas for $\int_I^{ba}, \dots, \int_{IV}^{ba}$.

    {\bf Computing integral I:} To start note that $\int_{I}^{ba}$ has exactly the same form as $\int_{II}$, \cref{eq:int_II_ba} except the roles of $\dot a_\F$ and $\dotblamF$ reversed. Since the latter is symmetric in $a_\F$ and $\blamF$. The form of $\int_{II}$ is
    $$
    \int_0^1 \dot a_\F \dot a_F \dt = 
                1 + \frac{\pi^2}{2} \sum_{n=1}^{N} n^2  a_n^2\\
    $$
    so 
    \begin{equation}
        \int_{I}^{ba} 
            = \lambda \intzo \dot a_\F \dotblamF \dt
            = \lambda + \lambda \frac{\pi^2}{2} \sum_{n=1}^{N} n^2 a_n b_n 
    \end{equation}

    \vskip 8pt

    {\bf Computing integral II:} Next $\int_{II}^{ba}$ involves $\dotblamF \dotblamF$ which is the same as $\int_I$, \cref{eq:int_I_ba} except for switching $\blamF$ for $a_\F$ and a factor of $\lambda^2$. The form of $\int_I$ is given by:
    $$
    \int_0^1 \dot a_\F \dot a_F \dt = 1 + \frac{\pi^2}{2} \sum_{n=1}^{N} n^2  a_n^2
    $$
    so
    \begin{equation}
        \int_{II}^{ba} 
            = \lambda^2 \intzo \dotblamF \dotblamF \dt + 
            = \lambda^2\left(1 + \frac{\pi^2}{2} \sum_{n=1}^{N} n^2  b_n^2\right)
    \end{equation}
    \vskip 8pt

    {\bf Computing integral III:} Next $\int_{III}^{ba}$ involves $a_\F \dotblamF$ which is the same as $\int_{IV}$, \cref{eq:int_IV_ba}, except for switching the roles of $\blamF$ for $a_\F$. The form of $\int_{IV}$ is:
    $$
     \kappa \lambda \int_0^1  \blamF \dot a_\F \dt =
                \kappa\lambda \left( \frac{1}{2} + \sumodd \frac{2(b_n- a_n)}{n \pi} + 2 \bigbs\sumnmodd a_n b_m  n \cdot \frac{m}{m^2 - n^2} \right)
    $$
    And therefore:

    \begin{equation}
        \int_{III}^{ba} 
            = \kappa \lambda \intzo a_\F\dotblamF \dt  + 
            = \kappa\lambda  \left( \frac{1}{2} + \sumodd \frac{2(a_n- b_n)}{n \pi} + 2\!\!\! \sumnmodd b_n a_m  n \cdot \frac{m}{m^2 - n^2} \right)
    \end{equation}

    \vskip 8pt

    {\bf Computing integral IV:} Finally $\int_{IV}$ involves $\blamF$ and $\dotblamF$ which mirrors \cref{eq:int_III_ba} with $\blamF$ replacing $a_\F$ and adding a factor of $\lambda^2$. Therefore since $\int_{IV}$ has the form:
    $$
    \kappa \int_0^1 \,\dot a_\F a_\F \dt = 
        \frac{\kappa}{2} + 2 \kappa \bigbs\!\sumnmodd  a_n a_m n \frac{m }{m^2 - n^2}
    $$
    we have:
    \begin{equation}
        \int_{IV}^{ba} 
            = \kappa \lambda^2 \intzo \blamF\dotblamF \dt \\ 
            = \frac{\kappa\lambda^2}{2} + 2 \kappa\lambda^2 \bigbs\!\sumnmodd  b_n b_m n \frac{m }{m^2 - n^2}
    \end{equation}

    \vskip 8pt

    {\bf Putting it all together:} To compute $\CostFba$ we compute the sum $\int_I^{ba} + \int_{II}^{ba} +\int_{III}^{ba} + \int_{IV}^{ba}$:

    \begin{equation}
    \begin{split}
        \label{eq:fourier-cost-fn-ba-messy}
        \CostFba &=\lambda + \lambda \frac{\pi^2}{2} \sum_{n=1}^{N} n^2 a_n b_n +\\
            &\quad  \lambda^2\left(1 + \frac{\pi^2}{2} \sum_{n=1}^{N} n^2  b_n^2\right) + \\
            &\quad  \kappa\lambda  \left( \frac{1}{2} + \sumodd \frac{2(a_n- b_n)}{n \pi} + 2\!\!\! \sumnmodd b_n a_m  n \cdot \frac{m}{m^2 - n^2} \right)+\\
            &\quad \frac{\kappa\lambda^2}{2} + 2 \kappa\lambda^2 \bigbs\!\sumnmodd  b_n b_m n \frac{m }{m^2 - n^2}
    \end{split}
    \end{equation}

    Collecting constants:

    \begin{equation}
    \begin{split}
        \label{eq:fourier-cost-fn-ba-messy-1}
        \CostFba &=\lambda + \lambda^2 + \frac{\kappa\lambda}{2}+\frac{\kappa\lambda^2}{2} + \\
            &\quad \lambda \frac{\pi^2}{2} \sum_{n=1}^{N} n^2 a_n b_n + 
                \lambda^2\frac{\pi^2}{2} \sum_{n=1}^{N} n^2  b_n^2 + \\
            &\quad  \kappa \lambda \left(\sumodd \frac{2(a_n- b_n)}{n \pi} +
                2\!\!\! \sumnmodd b_n a_m  n \cdot \frac{m}{m^2 - n^2} \right) + \\
            &\quad 2 \kappa\lambda^2 \bigbs\!\sumnmodd  b_n b_m n \frac{m }{m^2 - n^2}
    \end{split}
    \end{equation}

    Organizing:

    \begin{equation}
    \begin{split}
        \label{eq:fourier-cost-fn-ba-messy-2}
        \CostFba &=\lambda + \lambda^2 + \frac{\kappa\lambda}{2} +\frac{\kappa\lambda^2}{2} + \lambda\frac{\pi^2}{2} \sum_{n=1}^{N}  n^2\left(a_n b_n + \lambda b_n^2\right) + \kappa \lambda \sumodd \frac{2(a_n- b_n)}{n \pi} + \\
            &\quad 2\kappa\lambda\!\!\! \sumnmodd b_n a_m  n \cdot \frac{m}{m^2 - n^2}  +
                     2 \kappa\lambda^2 \bigbs\!\sumnmodd  b_n b_m n \frac{m }{m^2 - n^2}
    \end{split}
    \end{equation}

    Simplifying
    
    \begin{equation}
    \begin{split}
        \label{eq:fourier-cost-fn-ba-messy-3}
        \CostFba 
            &=\lambda + \lambda^2 + \frac{\kappa\lambda}{2} +\frac{\kappa\lambda^2}{2} + \lambda \frac{\pi^2}{2} \sum_{n=1}^{N}  n^2\left(a_n b_n + \lambda b_n^2\right) + 2\kappa \lambda \sumodd \frac{a_n- b_n}{n \pi} + \\
            &\quad 2\kappa\lambda\left(\sumnmodd b_n a_m  n \cdot \frac{m}{m^2 - n^2}  +
                     \lambda \bigbs\!\sumnmodd  b_n b_m n \frac{m }{m^2 - n^2}\right)
    \end{split}
    \end{equation}

    finally noting that the last is zero due to cancellations, we have:
    the last expression in \cref{eq:fourier-cost-fn-ba-messy-3} becomes

    \begin{equation}
            2\kappa\lambda \sumnmodd b_n a_m n\cdot \frac{m}{m^2 - n^2}
    \end{equation}

    Next we note:

    \begin{equation}
        \begin{split}
            \lambda + \lambda^2 + \frac{\kappa\lambda}{2} 
                &= \lambda\left(1 + \lambda + + \frac{\kappa}{2} + \frac{\kappa\lambda}{2}\right)\\     
                &= \lambda \frac{2(1+\lambda) + \kappa(1+\lambda)}{2}\\
                &= \frac{\lambda (1+\lambda )(2+\kappa)}{2}
        \end{split}
    \end{equation}

    so then

    \begin{equation}
        \CostFba = \frac{\lambda (1+\lambda)(2+\kappa)}{2} +
            \lambda \frac{\pi^2}{2} \sum_{n=1}^{N}  n^2\left(a_n b_n + \lambda b_n^2\right) +
             2\kappa \lambda \sumodd \frac{(a_n- b_n)}{n \pi} + 2\kappa\lambda \bigbs \sumnmodd  \frac{a_m b_n  n m}{m^2 - n^2}
    \end{equation}

    The conclusion after some re-arrangement and noting that the last term with $\lambda b_n b_m n m$ cancel so that we arrive at the formula conform to $\CostFab$ in Proposition \ref{prop:approx-cost-ba}:
    
    \begin{equation}
        \label{eq:fourier-cost-fn-ba-final-2}
        \CostFba = \lambda \left\{\frac{1}{2}(2+\kappa)(1+\lambda) \!+\!
            \frac{\pi^2}{2} \sum_{n=1}^{N}  \left(\lambda b_n^2 + a_n b_n\right)n^2 \!+\!
             \frac{2\kappa}{\pi} \sumodd \frac{a_n- b_n}{n} + 
             2\kappa\bigbs \sumnmodd  \frac{a_m b_n n m}{m^2 - n^2} \right\}
    \end{equation}

    \subsection{Key trigonometric identities}
    \label{sec:key-trig-identities}
    
    Finally we provide the key trigonometric identities we will use in this paper when dealing with Fourier series of trading strategies.

    \begin{subequations}        
    \begin{align}
        \int_0^1 \sin(n \pi t) \dt &=
            \begin{cases}
                \frac{2}{n \pi} & \text{if $n$ is odd} \\
                0 & \text{otherwise} 
            \end{cases} \label{eq:key-trig-a} \\
        \int_0^1 \cos(n \pi t) \dt &=
            \begin{cases}
            1 & \text{if } n = 0 \\
            0 & \text{if } n \neq 0
            \end{cases}  \label{eq:key-trig-b} \\
        \int_0^1 t \cos(n \pi t) \dt &=
            \begin{cases}
            -\frac{2}{n^2 \pi^2} & \text{if $n$ is odd}  \\
            0 & \text{otherwise}
            \end{cases} \label{eq:key-trig-c} \\
        \int_0^1 \cos(n \pi t) \cos(m \pi t) \dt &=
            \begin{cases}
                \frac{1}{2} & \text{if } n = m \\
                0 & \text{if } n \neq m\\
            \end{cases} \label{eq:key-trig-d} \\
        %
        \int_0^1 \cos(n \pi t) \sin(m \pi t) \, dt &= 
            \begin{cases}
                \frac{2 m }{\pi \left( m^2 - n^2 \right)} & \text{if $n+m$ is odd} \\
                0 & \text{otherwise}
            \end{cases}         \label{eq:int_cos_sin}
    \end{align}
    \end{subequations}

\section{Acknowledgements}

    I extend my sincere thanks to Vladimir Ragulin of Manifold Insights for building the  quadratic optimization code for solving the constrained and unconstrained optimizations, finding two-trader equilibria and producing the plots in this paper. He also introduced the use of state-space diagrams which have turned out to shed additional light on the nature of equilibria. His careful reading of drafts of this paper and transcription to code identified many extremely valuable insights and improvements in the work.

\bibliographystyle{unsrtnat}
\bibliography{references}

\end{document}